\documentclass[apj,numberedappendix]{emulateapj}

\usepackage{amsmath}
\usepackage[alpha,z,expert]{tmsfnts}
\usepackage{graphicx}

\newcommand{\Msun}{\ensuremath{M_\odot}}

\def\fgas{\ensuremath{f_{\text{gas}}}}
\def\Tspec{\ensuremath{T_{\text{spec}}}}
\def\Tmg{\ensuremath{T_{mg}}}
\def\rhogas{\ensuremath{\rho_{\text{gas}}}}
\def\rhotot{\ensuremath{\rho_{\text{tot}}}}

\def\Msun{\ensuremath{M_\odot}}

\def\mcc#1{\multicolumn{1}{c}{#1}}
\def\pz{\phantom{0}}

\begin{document}

\setcounter{dbltopnumber}{4}
\setcounter{totalnumber}{4}

\shorttitle{CLUSTER MASS PROFILES}
\shortauthors{VIKHLININ ET AL.}
\slugcomment{Submitted to ApJ July 2, 2005; astro-ph/0507092}

\title{\emph{Chandra} sample of nearby relaxed galaxy clusters: mass,
  gas fraction, and mass-temperature relation }

\author{%
A.~Vikhlinin\altaffilmark{1}\altaffilmark{,2},
A.~Kra\kern-0.05emvtsov\altaffilmark{3}, 
%A.~Kravtsov\altaffilmark{3}, % this is when using expert fonts
W. Forman\altaffilmark{1},
C. Jones\altaffilmark{1},
M. Markevitch\altaffilmark{1}\altaffilmark{,2},
S. S. Murray\altaffilmark{1},
L.~Van~Speybroeck\altaffilmark{1}\altaffilmark{,4}
}

\altaffiltext{1}{Harvard-Smithsonian Center for Astrophysics, 60 Garden
  St., Cambridge, MA 02138; avikhlinin@cfa.harvard.edu}
\altaffiltext{2}{Space Research Institute, Profsoyuznaya 84/32, Moscow,
  Russia.}  \altaffiltext{3}{Dept.\ of Astronomy and Astrophysics,
  Kavli Institute for Cosmological Physics, Enrico Fermi Institute,
  University of Chicago, Chicago, IL 60637} \altaffiltext{4}{This paper
  heavily uses \emph{Chandra} data of our late colleague.}

\begin{abstract}

  We present gas and total mass profiles for 13 low-redshift, relaxed
  clusters spanning a temperature range \mbox{0.7--9~keV}, derived from
  all available \emph{Chandra} data of sufficient quality. In all
  clusters, gas temperature profiles are measured to large radii
  (Vikhlinin et al.)  so that direct hydrostatic mass estimates are
  possible to nearly $r_{500}$ or beyond. The gas density was accurately
  traced to larger radii; its profile is not described well by a
  beta-model, showing continuous steepening with radius. The derived
  \rhotot{} profiles and their scaling with mass generally follow the
  Navarro-Frenk-White model with concentration expected for dark matter
  halos in $\Lambda$CDM cosmology. However, in three cool clusters, we
  detect a central mass component in excess of the NFW profile,
  apparently associated with their cD galaxies. In the inner region
  ($r<0.1~r_{500}$), the gas density and temperature profiles exhibit
  significant scatter and trends with mass, but they become nearly
  self-similar at larger radii.  Correspondingly, we find that the slope
  of the mass-temperature relation for these relaxed clusters is in good
  agreement with the simple self-similar behavior, $M_{500}\propto
  T^\alpha$ where $\alpha=(1.5-1.6)\pm0.1$, if the gas temperatures are
  measured excluding the central cool cores.  The normalization of this
  $M-T$ relation is significantly, by $\approx 30\%$, higher than most
  previous X-ray determinations. We derive accurate gas mass fraction
  profiles, which show increase both with radius and cluster mass.  The
  enclosed \fgas{} profiles within $r_{2500}\simeq0.4\,r_{500}$ have not
  yet reached any asymptotic value and are still far (by a factor of
  $1.5-2$) from the Universal baryon fraction according to the CMB
  observations. The \fgas{} trends become weaker and its values closer
  to Universal at larger radii, in particular, in spherical shells
  $r_{2500}<r<r_{500}$.

\end{abstract}


\section{Introduction}

Observations of galaxy clusters offer a number of well-established
cosmological tests (see \citealt{2005RvMP...77..207V} for a recent
review). 
% The number density of clusters as a function
% of mass is directly linked to the power spectrum of the matter density
% perturbations \citep{1974ApJ...187..425P}. The fraction of the baryonic
% matter in the cluster total mass approximates that for the entire
% Universe and thus independently constrains the ratio $\Omega_b/\Omega_m$
% \citep{1993Natur.366..429W}. The apparent redshift dependence of the
% cluster baryon fraction is, in principle, an attractive indicator of the
% angular size distance
% \citep{1996PASJ...48L.119S,1997NewA....2..309P,2004MNRAS.353..457A}. 
% Another, classical test for the angular size distance uses the
% Sunyaev-Zeldovich effect \citep{1972CoASP...4..173S}. 
Many of these tests rely on the paradigm in which clusters are composed
mostly of collisionless cold dark matter (CDM), and virialized objects
form from scale-free, Gaussian initial density perturbations.  Numerical
simulations of cluster formation in CDM cosmology are used to calibrate
essential theoretical ingredients for cosmological tests, such as the
detailed shape of mass function models
\citep{1999MNRAS.308..119S,2001MNRAS.321..372J} or the average baryon
bias within clusters.  The CDM paradigm and numerical simulations make
clear predictions for the structure of clusters, for example that they
should have a universal density profile
\citep{dubinski_carlberg91,navarro_etal96}, that their observable
properties should exhibit scaling relations, and so on. Confrontation of
these predictions with the results of high-quality observations is a
necessary consistency check. Any significant disagreement, beyond that
attributable to variations of an underlying cosmology, indicates either
that the theoretical models are not sufficiently accurate (e.g., they do
not include important non-gravitational processes) or that there are
significant hidden biases in the observational studies. Both
possibilities are red flags for the application of cluster-based
cosmological tests in the present ``era of precision cosmology''. 

The above underscores the need for high-quality observational studies of
representative cluster samples. Of particular importance are
measurements of the distribution of the dark matter and the dominant
baryonic component, the hot intracluster medium (ICM). The mass
distribution in dynamically relaxed clusters can be reconstructed using
several approaches, of which the X-ray method is one of the most widely
used (a recent review of the mass determination techniques can be found,
e.g., in \citealt{2005RvMP...77..207V}). X-ray telescopes directly map
the distribution of the ICM. The ICM in relaxed clusters should be close
to hydrostatic equilibrium and then the spatially-resolved X-ray
spectral data can be used to derive the total mass as a function of
radius \citep[e.g.,][]{1978ApJ...219..413M,sarazin88}. 

Using X-ray cluster observations for cosmological applications has a
long history. Early determinations of the amplitude of density
fluctuations can be found in \cite{1990ApJ...351...10F},
\cite{1991ApJ...372..410H}, \cite{1992ApJ...386L..33L}, and
\cite{1993MNRAS.262.1023W}.  \citeauthor{1991ApJ...372..410H} also used
the shape of the cluster temperature function to constrain the slope of
the perturbation spectrum on cluster scales. 
\citet{1992A&A...262L..21O} proposed that $\Omega_m$ can be constrained
by evolution of the temperature function, and this test was applied to
$z\approx 0.3-0.4$ clusters by \citet{1997ApJ...489L...1H} and
\citet{1998MNRAS.298.1145E}. Along a different line of argument,
\citet{1993Natur.366..429W} obtained the first determination of
$\Omega_m$, assuming that the baryon fraction in clusters approximates
the cosmic mean. Attempts to use the gas fraction as a distance
indicator were made in papers by \citet{1999ApJ...517...70R} and
\citet{1999MNRAS.305..834E}.  The common limitation of these early
studies is the rather uncertain estimates of the cluster total mass. 
Accurate cluster mass measurements at large radii are challenging with
any technique. The X-ray method, for example, requires that the ICM
temperature be measured locally in the outer regions; the hydrostatic
mass estimate is only as accurate as $T$ and $dT/dr$ at radius $r$. 

Spatially resolved cluster temperature measurements first became
possible with the launch of \emph{ASCA} and \emph{Beppo-SAX}
\citep[e.g.,][]{1998ApJ...503...77M,2002ApJ...567..163D}. Truly accurate
temperature profiles are now provided by \emph{Chandra} and
\emph{XMM-Newton}. \emph{Chandra} results on the mass distribution in
the innermost cluster regions were reported by
\citet{2001ApJ...557..546D}, \citet{2003ApJ...586..135L},
\citet{2004ApJ...604..116B}, and \citet{2004ApJ...617..303A}, and within
$\sim 1/4$ of $r_{200}$ in a series of papers by \citet[][and references
therein]{2004MNRAS.353..457A}. The \emph{XMM-Newton} mass measurements
at large radii ($\sim 0.5\,r_{200}$) in 10 low-redshift clusters were
recently published by \citet{2005A&A...435....1P} and
\citet{2005astro.ph..2210A}. 

\begin{deluxetable}{lcrrc}
\tablecaption{Cluster Sample\label{tab:sample}}
\tablehead{
\colhead{Cluster\hspace*{20mm}} &
\colhead{$z$} &
\colhead{$r_{\text{min}}$\rlap{\tablenotemark{a}}} &
\colhead{$r_{\text{det}}$\rlap{\tablenotemark{b}}} &
\colhead{\emph{ROSAT}\rlap{\tablenotemark{c}}}
}
\startdata
A133\dotfill           & 0.0569 &  40 & 1100 & +      \\
A262\dotfill           & 0.0162 &  10 &  450 & +      \\
A383\dotfill           & 0.1883 &  25 &  800 & \nodata\\
A478\dotfill           & 0.0881 &  30 & 2000 & +      \\
A907\dotfill           & 0.1603 &  40 & 1300 & \nodata\\
A1413\dotfill          & 0.1429 &  20 & 1800 & \nodata\\
A1795\dotfill          & 0.0622 &  40 & 1500 & +      \\
A1991\dotfill          & 0.0592 &  10 & 1000 & +      \\
A2029\dotfill          & 0.0779 &  20 & 2250 & +      \\
A2390\dotfill          & 0.2302 &  80 & 2500 & \nodata\\
RXJ\,1159+5531\dotfill & 0.0810 &  10 &  600 & \nodata\\
MKW4\dotfill           & 0.0199 &   5 &  550 & +      \\
USGC~S152\dotfill      & 0.0153 &  20 &  300 & \nodata\\[-7pt]
\enddata
\tablenotetext{a}{--- Inner boundary (kpc) of the radial range used for
  the temperature profile fit (\S\,\ref{sec:results:indiv}).} 
\tablenotetext{b}{--- The radius (kpc) where X-ray brightness is
  detected at \mbox{$>3\sigma$}, or the outer boundary of the \emph{Chandra}
  field of view.} 
\tablenotetext{c}{--- Those clusters for which we use also \emph{ROSAT}
  PSPC surface brightness measurements. 
  \vspace*{-1.0\baselineskip}
} 
\end{deluxetable}

\defcitealias{2004astro.ph.12306V}{Paper~I} In this paper, we present
the mass measurements in a sample of 13 low-redshift clusters whose
temperature profiles were derived in \citep[Paper~I hereafter;
Table~\ref{tab:sample}]{2004astro.ph.12306V}. These clusters were
observed with sufficiently long \emph{Chandra} exposures that the
temperature profiles can be measured to $0.75\,r_{500}$ ($\sim
0.5\,r_{200}$) in all objects and in 5 cases, can be extended outside
$r_{500}$. All these objects have a very regular X-ray morphology and
show only weak signs of dynamical activity, if any.  Even though the
present sample is not a statistically complete snapshot of the cluster
population, it represents an essential step towards reliable
measurements of the cluster properties to a large fraction of the virial
radius. 

The Paper is organized as follows. Our approach to 3-dimensional
modeling of the observed projected quantities is described in
\S\,\ref{sec:prof:modeling}. Results for individual clusters are
presented in \S\,\ref{sec:results:indiv} and self-similarity of their
temperature and density profiles is discussed in \S\,\ref{sec:T:prof}
and \ref{sec:scaled:prof}. Our data lead to accurate determination of
the $M-T$ relation for relaxed clusters (\S\,\ref{sec:MT}). In
\S\,\ref{sec:fgas}, we discuss the observed systematic variations of ICM
mass fraction with radius and cluster mass. 

To compute all distance-dependent quantities, we assume $\Omega_M=0.3$,
$\Lambda=0.7$, $h=0.71$. Uncertainties are quoted at 68\% CL.  Cluster
masses are determined at radii $r_{500}$ and $r_{2500}$, corresponding
to overdensities 500 and 2500 relative to the critical density at the
cluster redshift.

\section{X-ray Data Analysis}
\label{sec:data}

The main observational ingredients for the present analysis are radial
profiles of the projected temperature and X-ray surface brightness. We
refer the reader to \citetalias{2004astro.ph.12306V} for an extensive
description of all technical aspects of the \emph{Chandra} data
reduction and spectral analysis and discuss here only the X-ray surface
brightness profile measurements. 

\begin{figure}
\centerline{\includegraphics[width=0.95\linewidth]{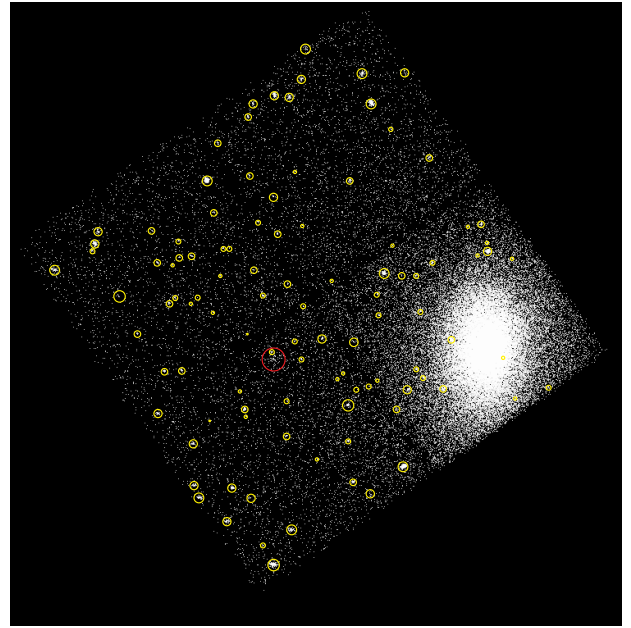}}
\caption{Detected sources in the \emph{Chandra} ACIS-I field of A1413. 
  Yellow circles mark point sources. The only detectable extended X-ray
  source (other than A1413) is marked by the red circle. ACIS-I field of
  view is $16'\times16'$.}\label{fig:a1413src}
\end{figure}

First, we detected and masked out small-scale X-ray sources detectable
in either soft (0.7--2~keV) or hard (2--7~keV) energy bands. Detection
was performed using the wavelet decomposition algorithm described in
\cite{1998ApJ...502..558V}.  Detection thresholds were chosen to allow
1--2 false sources per field of view. The results were hand-checked for
each cluster and exclusion radii adjusted if needed.  Source detection
for one of our clusters (Fig.\,\ref{fig:a1413src}) is illustrated in
Fig.\,\ref{fig:a1413src}. Note that in addition to point sources, we
detected and excluded small-scale extended X-ray sources (a typical
example of a source of this type is shown by the red circle in
Fig.\,\ref{fig:a1413src}). This should remove at least the brightest of
the cold gas clumps associated with groups and individual galaxies which
are present in the cluster volume
\citep{2004ApJ...606..635M,2003ApJ...587..524N,2004ApJ...606L..97D} and
could bias measurements of the global cluster parameters
\citep{2005ApJ...618L...1R}. The typical limiting flux for detection of
compact extended sources is $\sim
3\times10^{-15}$~erg~s$^{-1}$~cm$^{-2}$ in the 0.5--2~keV band; this
corresponds to a luminosity of $\sim 1.5\times10^{42}$~erg~s$^{-1}$ for
the median redshift of our sample, $z=0.06$. 
%
% 30cnt/50ksec*3.5412E-13/0.7497E-01 = 
%

The surface brightness profiles are measured in the 0.7--2~keV energy
band, which provides an optimal ratio of the cluster and background flux
in \emph{Chandra} data.  The blank-field background is subtracted from
the cluster images, and the result is flat-fielded using exposure maps
that include corrections for CCD gaps and bad pixel, but do not include
any spatial variations of the effective area.  We then subtract any
small uniform component corresponding to adjustments to the soft X-ray
foreground that may be required (see \citetalias{2004astro.ph.12306V}
for details of background modeling).  This correction is done separately
for the back-illuminated (BI) and front-illuminated (FI) CCDs because
they have very different low-energy effective area. 

We then extracted the surface brightness profiles in narrow concentric
annuli ($r_{\text{out}}/r_{\text{in}}=1.05$) centered on the cluster
X-ray peak and computed the area-averaged \emph{Chandra} effective area
for each annulus (see Paper~I for details on calculating the effective
area). Using the effective area and observed projected temperature and
metallicity as a function of radius, we converted \emph{Chandra} count
rate in the 0.7--2~keV band into integrated emission measure,
$\mathit{EM}=\int n_e \,n_p\,dV$, within the cylindrical shell. We have
verified that these ``physical'' cluster brightness profiles derived
from BI and FI CCDs are always in excellent agreement in the overlapping
radial range.  The profiles from both CCD sets and different pointings
were combined using the statistically optimal weighting.  Poisson
uncertainties of the original X-ray data were propagated throughout this
procedure. 

For lower-redshift clusters in our sample, the statistical accuracy of
the surface brightness at large radii is limited mostly by the
\emph{Chandra} field of view. The analysis in such cases can benefit
from also using data from the \emph{ROSAT} PSPC pointed observations. We
used flat-fielded \emph{ROSAT} images in the 0.7--2~keV band produced
using S.~Snowden's software \citep{1994ApJ...424..714S} and reduced as
described in \cite{1999ApJ...525...47V}. \emph{ROSAT} surface brightness
profiles were converted to projected emission measure and were used as a
second independent dataset in modeling the gas density distribution. We
did not use the \emph{ROSAT} data in the central $2'$ from the cluster
center, because this region can be affected by the PSPC angular
resolution ($\sim 25''$ FWHM). At larger radii, we always find excellent
agreement between \emph{Chandra} and \emph{ROSAT} PSPC surface
brightness data (an example is shown in Fig.\,\ref{fig:sxfit}). Below,
we used the combined \emph{Chandra} and \emph{ROSAT} analysis for those
clusters with sufficiently deep PSPC exposures --- A133, A262, A478,
A1795, A1991, A2029, and MKW4.

\section{Modeling of temperature and surface brightness profiles}
\label{sec:prof:modeling}

We model the observed X-ray surface brightness and projected temperature
profiles using the following general approach. The 3D profiles of gas
density and temperature are represented with analytic functions which
are smooth, but have freedom to describe a wide range of the possible
profiles. The models are projected along the line of sight and fit to
the data. The best fit 3D model is used to derive all interesting
cluster parameters, such as the total gravitating mass.  Measurement
uncertainties for all quantities are estimated using Monte-Carlo
simulations in which random statistical errors are added to the data and
the full analysis is repeated using these simulated data as an input. 

Reliability of this modeling approach was tested by applying our
\emph{Chandra} data analysis procedures to ``observations'' of clusters
from high-resolution numerical simulations (Nagai et al., in
preparation).  This analysis demonstrated that the three-dimensional gas
density and temperature profiles of relaxed clusters are reconstructed
within a few percent. 

\begin{deluxetable*}{lccrrccccccc}
\tablecaption{Best Fit Parameters for Gas Density Profiles
  \label{tab:rhogas:fit}}
\tablehead{
\colhead{Cluster\hspace*{20mm}} &
\colhead{$r_{\text{det}}$\rlap{\tablenotemark{a}}} &
\colhead{$n_0$} &
\colhead{$r_c$} &
\colhead{$r_s$} &
\colhead{$\alpha$} &
\colhead{$\beta$} &
\colhead{$\varepsilon$} &
\colhead{$n_{02}$} &
\colhead{$r_{c2}$} &
\colhead{$\beta_2$} &
\colhead{\emph{ROSAT}\rlap{\tablenotemark{b}}}
\\
 & \colhead{(kpc)} & \colhead{$10^{-3}$~cm$^{-3}$}
 & \colhead{(kpc)} & \colhead{(kpc)} & & & & \colhead{$10^{-1}$~cm$^{-1}$}
}
\startdata
A133          \dotfill & 1100& \pz 2.968 & 142.7 & 1423.3 & 0.996 & 0.575 & 5.000 &   0.276 &      33.44 &   0.980 & +      \\
A262          \dotfill &  450& \pz 3.434 &  45.2 &  350.8 & 1.674 & 0.333 & 1.806 & \nodata &    \nodata & \nodata & +      \\
A383          \dotfill &  800& \pz 7.000 & 115.2 &  422.3 & 2.018 & 0.583 & 0.740 &   1.014 &   \pz 0.08 &   1.000 & \nodata\\
A478          \dotfill & 2000& \pz 8.169 & 177.1 & 3148.2 & 1.493 & 0.715 & 5.000 &   0.584 &      24.00 &   1.000 & +      \\
A907          \dotfill & 1300& \pz 6.257 & 136.9 & 1885.1 & 1.554 & 0.594 & 4.986 & \nodata &    \nodata & \nodata & \nodata\\
A1413         \dotfill & 1800& \pz 5.526 & 186.3 & 2077.1 & 1.217 & 0.651 & 4.991 & \nodata &    \nodata & \nodata & \nodata\\
A1795         \dotfill & 1500&    14.993 &  72.8 & 1030.8 & 1.060 & 0.545 & 3.474 & \nodata &    \nodata & \nodata & +      \\
A1991         \dotfill & 1000& \pz 9.373 &  44.4 &  998.2 & 1.516 & 0.501 & 5.000 &   0.999 &   \pz 5.00 &   1.165 & +      \\
A2029         \dotfill & 2250&    16.469 &  80.7 &  870.6 & 1.131 & 0.539 & 1.650 &   3.741 &   \pz 5.00 &   1.000 & +      \\
A2390         \dotfill & 2500& \pz 3.069 & 353.6 & 1200.0 & 1.917 & 0.696 & 0.240 & \nodata &    \nodata & \nodata & \nodata\\
RXJ\,1159+5531\dotfill &  600& \pz 0.198 & 613.8 &  961.5 & 1.762 & 1.215 & 4.939 &   0.416 &      12.66 &   1.000 & \nodata\\
MKW4          \dotfill &  550& \pz 0.280 & 488.6 & 1081.6 & 1.628 & 1.224 & 0.000 &   0.189 &      11.08 &   0.661 & +      \\
USGC~S152     \dotfill &  300&    17.450 &   8.1 &  467.5 & 2.644 & 0.453 & 3.280 & \nodata &    \nodata & \nodata & \nodata
\enddata
\tablenotetext{a}{--- The radius (kpc) where X-ray brightness is
  detected at \mbox{$>3\sigma$}, or the outer boundary of the \emph{Chandra}
  field of view, whichever is smaller.} 
\tablenotetext{b}{--- Those clusters for which we use also \emph{ROSAT}
  PSPC surface brightness measurements.
\vspace*{-1\baselineskip}
} 
\tablecomments{Derived densities and radii scale with the Hubble
  constant as $h^{1/2}$ and $h^{-1}$, respectively.
} 
\end{deluxetable*}

\subsection{Gas Density Model}
\label{sec:gas:density:model}

\begin{figure}
\vspace*{-2\baselineskip}
\includegraphics[width=0.99\linewidth]{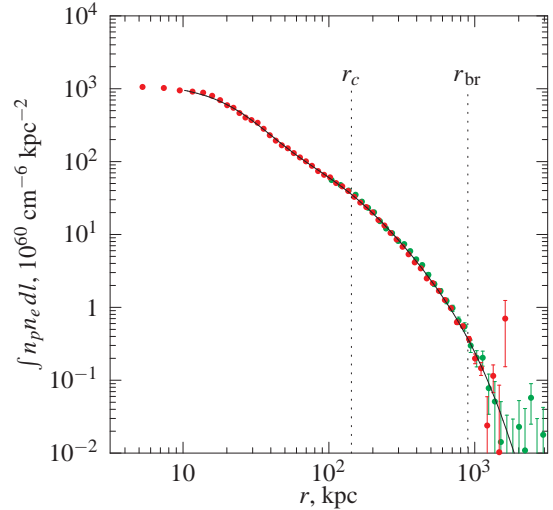}
\vspace*{-0.5\baselineskip}
\caption{Observed projected emissivity profile for A133. \emph{Chandra}
  and \emph{ROSAT} PSPC data are shown in red and green, respectively. 
  Solid line shows the best fit to the 3D gas density model
  (eq.\,[\ref{eq:density:model}]). The slope of the emissivity profile
  steepens by $1$ at radius $r_{\text{br}} = r_s
  (\varepsilon-1)^{-1/\gamma}$.} 
\label{fig:sxfit}
\end{figure}

The analytic expression we use for the 3D gas density distribution is
obtained by modifying the traditionally used $\beta$-model
\citep{1978A&A....70..677C}. Modifications are designed to represent
essential features of the observed X-ray surface brightness profiles. 
Gas density in the centers of relaxed clusters, such as ours,
usually has a power law-type cusp instead of a flat core and so the
first modification is 
\begin{equation}
  \frac{n_0^2}{(1+r^2/r_c^2)^{3\beta}} \quad \longrightarrow \quad
  \frac{n_0^2\;(r/r_c)^{-\alpha}}{(1+r^2/r_c^2)^{3\beta-\alpha/2}}. 
\end{equation}
This model was also used by, e.g., \cite{2004A&A...423...33P}. 

At large radii, the observed X-ray brightness profiles often steepen at
$r>0.3\,r_{200}$ relative to the power law extrapolated from smaller
radii \citep{1999ApJ...525...47V}. This change of slope can be modeled
as
\begin{equation}
  \frac{n_0^2}{(1+r^2/r_c^2)^{3\beta}} \quad \longrightarrow \quad
  \frac{n_0^2}{(1+r^2/r_c^2)^{3\beta}}\,\frac{1}{(1+r^\gamma/r_s{}^\gamma)^{\varepsilon/\gamma}},
\end{equation}
where the additional term describes a change of slope by $\varepsilon$
near the radius $r_s$ and the parameter $\gamma$ controls the width of
the transition region. Finally, we add a second $\beta$-model component
with small core-radius to increase modeling freedom near the cluster
centers. The complete expression for the emission measure profile is
\begin{equation}\label{eq:density:model}
  \begin{split}
    n_p\,n_e = n_0^2\;\frac{(r/r_c)^{-\alpha}}{(1+r^2/r_c^2)^{3\beta-\alpha/2}}
                \;&
                \frac{1}{(1+r^\gamma/r_s{}^\gamma)^{\varepsilon/\gamma}}\\
                &+ \frac{n_{02}^2}{(1+r^2/r_{c2}^2)^{3\beta_2}}. 
  \end{split}
\end{equation}

All our clusters can be fit adequately by this model with a fixed
$\gamma=3$. All other parameters were free. The only constraint we used
to exclude unphysically sharp density breaks was $\varepsilon<5$. The
analytic model [\ref{eq:density:model}] has great freedom and can fit
independently the inner and outer cluster regions.  This is important
for avoiding biases in the mass measurements at large radii and also for
more realistic uncertainty estimates. The best-fit model for the
observed surface brightness in A133 is shown by a solid line in
Fig.\,\ref{fig:sxfit}.

Parameters in eq.\,[\ref{eq:density:model}] are strongly correlated and
therefore their individual values are degenerate.  This is not a problem
because our goal is only to find a smooth analytic expression for the
gas density which is consistent with the observed X-ray surface
brightness throughout the radial range of interest. 

\subsection{Temperature Profile Model}

Several previous studies used a polytropic law, $T(r)\propto
[\rho_{\text{gas}}(r)]^{\gamma-1}$, to model non-constant cluster
temperature profiles
\citep{1999ApJ...527..545M,2001A&A...368..749F,2002A&A...394..375P}.  We
take a different approach because the polytropic model is in fact a poor
approximation of the  temperature profiles at large radii
\citep{1998ApJ...503...77M,2002ApJ...567..163D}, which is apparent in
our more accurate \emph{Chandra} measurements. 
% For example, the effective polytropic index, $\gamma=d
% \log T /d \log \rho + 1$, changes in one of our best-observed clusters,
% A133, from $\gamma=1.07$ at $r=200$~kpc to $1.27$ at $r=800$~kpc. 
% \***{Errors?} 

The projected temperature profiles for all our clusters show similar
behavior (see Fig.\,15 in \citetalias{2004astro.ph.12306V} and
Fig.\,\ref{fig:a133:res}--\ref{fig:s152:res} below). The temperature has
a broad peak near 0.1--0.2 of $r_{200}$ and decreases at larger radii in
a manner consistent with the results of earlier \emph{ASCA} and
\emph{Beppo-SAX} observations
\citep{1998ApJ...503...77M,2002ApJ...567..163D}, reaching approximately
50\% of the peak value near $0.5\,r_{200}$. There is also a temperature
decline towards the cluster center probably because of the presence of
radiative cooling. We construct an analytic model for the temperature
profile in 3D, in such a way that it can describe these general
features. Outside the central cooling region, the temperature profile
can be adequately represented as a broken power law with a transition
region:
\begin{equation}\label{eq:tprof:main}
  t(r) =\frac{(r/r_t)^{-a}}{(1+(r/r_t)^b)^{c/b}}. 
\end{equation}
The temperature decline in the central region in most clusters can be
described as
\begin{equation}\label{eq:tprof:cool}
  t_{\text{cool}}(r) = (x+T_{\text{min}}/T_0)/(x+1), \quad
  x=(r/r_{\text{cool}})^{a_{\text{cool}}}, 
\end{equation}
\citep*{2001MNRAS.328L..37A}. Our final model for the 3D temperature
profile is the product of eq.\,[\ref{eq:tprof:main}] and
[\ref{eq:tprof:cool}],
\begin{equation}\label{eq:tprof}
  T_{\mathrm{3D}}(r) =  T_0\times t_{\text{cool}}(r)\times t(r). 
\end{equation}
Our model has great functional freedom (9 free parameters) and can
adequately describe almost any type of smooth temperature distribution
in the radial range of interest. 

This model is projected along the line of sight to fit the observed
projected temperature profile. This projection requires proper weighting
of multiple temperature components. We use the algorithm described in
\cite{astro-ph/0504098} that very accurately predicts the
single-temperature fit to multi-component spectra over a wide range of
temperatures. The inputs for this algorithm are the 3D profiles of the
gas temperature, density, and metallicity. The only missing ingredient
in our case is the 3D metallicity profile. We use instead the projected
metallicity distributions presented in Paper~I. The difference between
the projected and 3D abundance profiles leads to very small corrections
in the calculation of the project temperature that are negligible for
our purposes. We also note that the 3D temperature fit at large radii is
rather insensitive to the choice of weighting algorithm.  For example,
we tested the commonly used method of weighting $T$ with the square of
the gas density and obtained very similar results.  The primary reason
for stability of the $T_{\text{proj}}$ calculations is that the ICM
emissivity is a strongly decreasing function of radius and most of the
emission observed at a projected distance $b$ comes from a narrow radial
range near $r=b$. 

In several cases, the cluster X-ray brightness is detected beyond the
outer boundary of the \emph{Chandra} temperature profile. Typical
examples are A262 where the \emph{ROSAT} PSPC profile extends to $50'$,
well beyond the \emph{Chandra} FOV, and A383 where the X-ray brightness
is detectable in the \emph{Chandra} image to 1300~kpc, while the
temperature profile is sufficiently accurate only within the central
750~kpc.  Detection of ICM emission at large radii sets a lower limit to
the temperature and thus provides additional information for the $T(r)$
modeling. We required that the 3D temperature model exceeds 0.5~keV at
$r=r_{\text{det}}$, the radius where the X-ray brightness is at least
$3\sigma$ significant ($r_{\text{det}}$ are reported in
Table~\ref{tab:sample}). 

We excluded from the fit the data within the inner cutoff radius,
$r_{\text{min}}$ (listed in Table~\ref{tab:sample}), which was chosen to
exclude the central temperature bin (10--20~kpc) because the ICM is
likely to be multi-phase at these small radii. In A133 and A478,
$r_{\text{min}}$ was increased to exclude substructures associated with
activity of the central AGN. The cutoff radius was increased also in
A1795, A2390, and USGC~S152, because our analytic model is a poor fit to
the inner temperature profile in these clusters. The choice of
$r_{\text{min}}$ is unimportant because we are primarily interested in
the cluster properties at large radii. 
\begin{figure*}
\vspace*{-1.75\baselineskip}
\centerline{\includegraphics[width=1.00\linewidth]{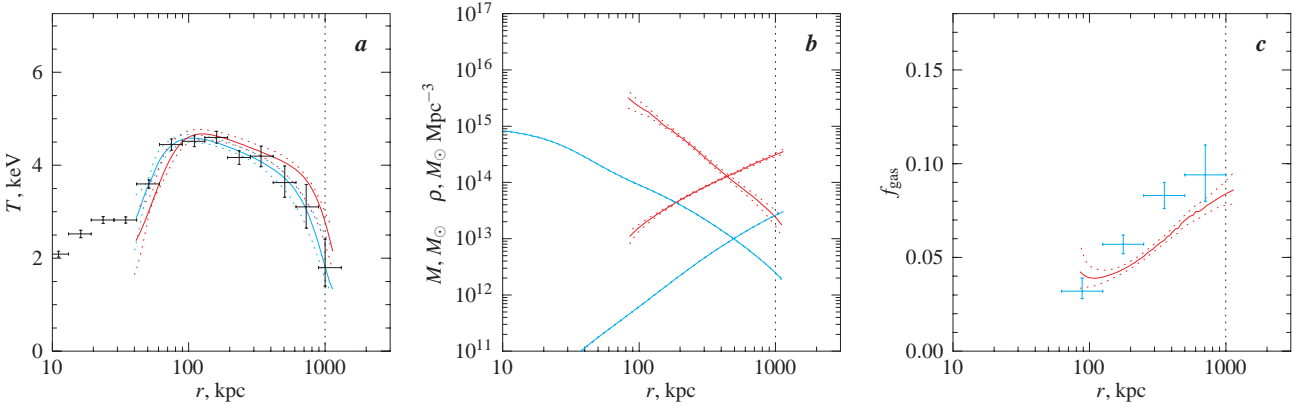}}
\vspace*{-0.5\baselineskip}
\caption{Results for A133. \emph{(a)} --- Temperature profile.  Observed
  projected temperatures are shown by crosses. Solid red and blue lines
  show the best fit 3D model and the corresponding projected profile,
  respectively. Dotted lines indicate the 68\% CL uncertainty interval
  obtained from Monte-Carlo simulations (see text).  Models are shown in
  the radial range $r_{\text{min}}-r_{\text{det}}$ (see text). 
  \emph{(b)} --- Mass and density profiles. $M(r)$ increases with radius
  and $\rho(r)$ decreases. Red and blue lines show results for the total
  mass and gas mass, respectively. \emph{(c)} --- Gas mass fraction as a
  function of radius. Lines show the enclosed
  $f_{\text{gas}}=M_{\text{gas}}(<r)/M_{\text{tot}}(<r)$. The local gas
  fraction ($\rho_{\text{gas}}/\rho_{\text{tot}}$) in the radial range
  directly covered by the \emph{Chandra} temperature data is shown by
  crosses. The vertical line shows the radius $r_{500}$ derived from the
  best fit mass model.} 
\label{fig:a133:res}
\end{figure*}

\begin{figure*}
\vspace*{-2.25\baselineskip}
\centerline{\includegraphics[width=1.00\linewidth]{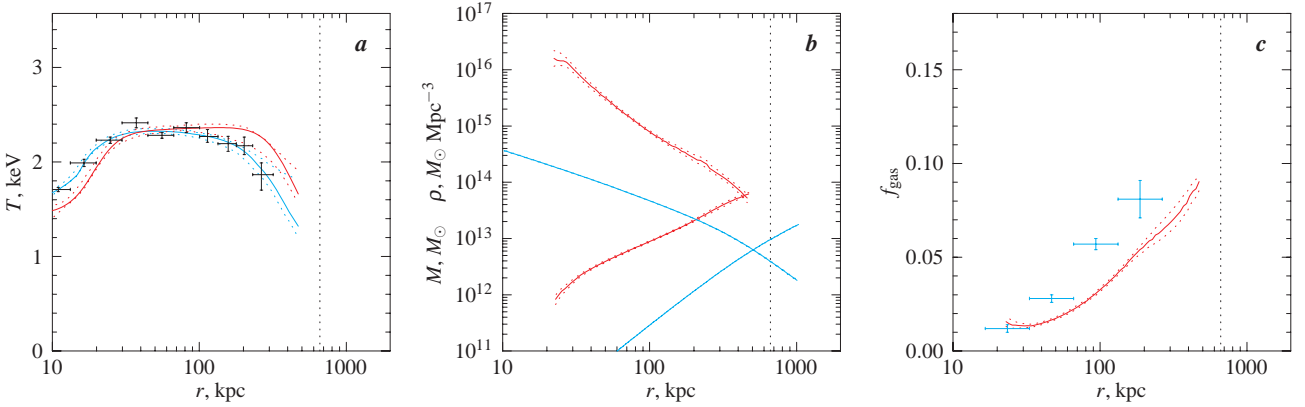}}
\vspace*{-0.5\baselineskip}
\caption{Results for A262. See caption to Fig.\,\ref{fig:a133:res}.} 
\end{figure*}

\subsection{Uncertainties}
\label{sec:uncert:Monte-Carlo}

Our analytic models for the gas density and temperature profiles have many
free parameters and strong intrinsic degeneracies between parameters.
Therefore, uncertainty intervals for all quantities of interest were
obtained from Monte-Carlo simulations.  The simulated data were realized by
scattering the observed brightness and temperature profiles according to a
Gaussian distribution with a dispersion equal to measurement uncertainties.
The surface brightness and temperature models were fit to the simulated data
and the full analysis repeated.  The uncertainty on all quantities of
interest were obtained by analyzing the distribution derived from 1000--4000
simulated profiles we generated for each cluster.

\subsection{Mass Derivation}
\label{sec:mass:derivation}

Given 3D models for the gas density and temperature profiles, the total
mass within the radius $r$ can be estimated from the hydrostatic
equilibrium equation \citep[e.g.,][]{sarazin88}:
\begin{equation}\label{eq:hydro}
  M(r) = -3.71\times10^{13}\,\Msun\,
  T(r)\;r
  \left(\frac{d\,\log \rho_g}{d\,\log r}+\frac{d\,\log T}{d\,\log r}\right),
\end{equation}
where $T$ is in units of keV and $r$ is in units of Mpc. Given $M(r)$,
we can calculate the total matter density profile, $\rho(r) = (4\pi
r^2)^{-1}\,dM/dr$. We also compute the total mass at several overdensity
levels, $\Delta$, relative to the critical density at the cluster
redshift, by solving equation
\begin{equation}\label{eq:overdensity}
  M_{\Delta}(r_\Delta) = \Delta\, 4/3\,\pi\,r_\Delta^3\,  \rho_c(z),
\end{equation}
(see, e.g., \citealt{2001A&A...367...27W} for discussion of different
cluster mass definitions). 

\begin{figure*}
\vspace*{-2.25\baselineskip}
\centerline{\includegraphics[width=1.00\linewidth]{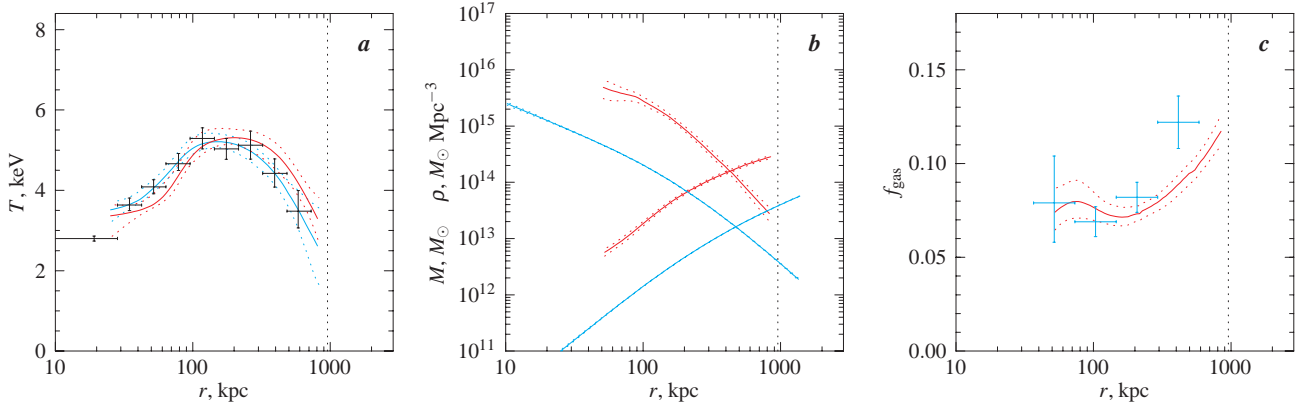}}
\vspace*{-0.5\baselineskip}
\caption{Results for A383. See caption for Fig.\,\ref{fig:a133:res}.} 
\end{figure*}

\begin{figure*}
\vspace*{-2.5\baselineskip}
\centerline{\includegraphics[width=1.00\linewidth]{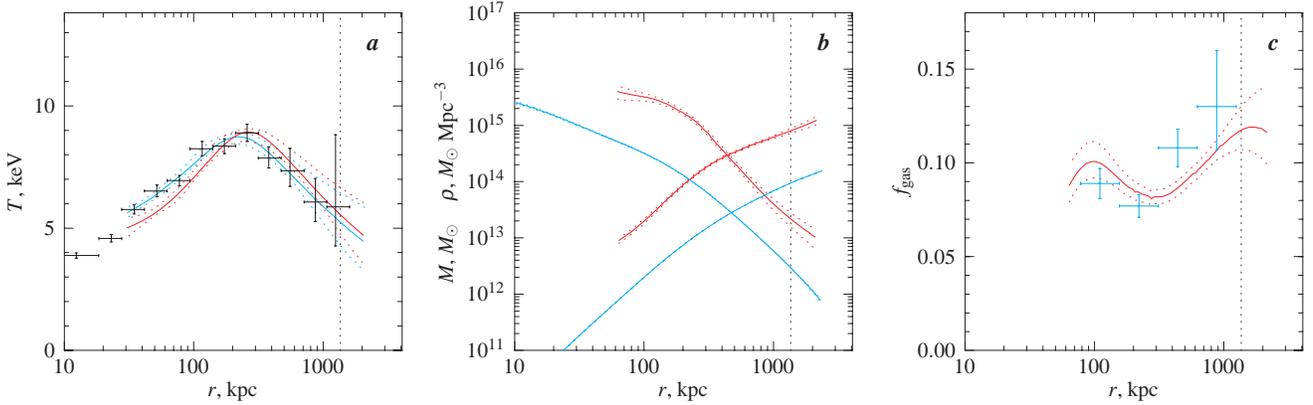}}
\vspace*{-0.5\baselineskip}
\caption{Results for A478. See caption for Fig.\,\ref{fig:a133:res}.} 
\end{figure*}

\begin{figure*}
\vspace*{-2.5\baselineskip}
\centerline{\includegraphics[width=1.00\linewidth]{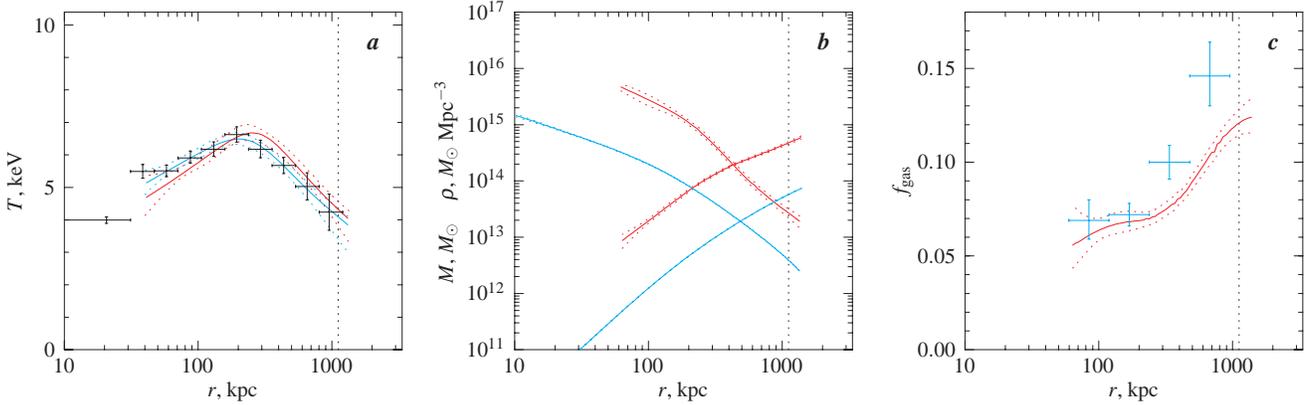}}
\vspace*{-0.5\baselineskip}
\caption{Results for A907. See caption for Fig.\,\ref{fig:a133:res}.} 
\end{figure*}

The ICM particle number density profile is given directly by the
analytic fit to the projected emission measure profile
(\S\,\ref{sec:gas:density:model}) and it is easily converted to the gas
density. For the cosmic plasma with the He abundance from
\cite{1989GeCoA..53..197A}, $\rho_g = 1.252\,m_p\,(n_p n_e)^{1/2}$. 

The total mass derived from eq.\,[\ref{eq:hydro}] is a complex
combination of parameters which define our $T(r)$ and $\rho_g(r)$
models.  Its uncertainties are best derived via Monte-Carlo simulations
as described in \S\,\ref{sec:uncert:Monte-Carlo}.  We used the best fit
$T(r)$ model obtained for each realization of the temperature profile to
compute all quantities of interest ($M_\Delta$, $M_g$, gas mass
fraction, etc.). The peak in the obtained distribution corresponds the
most probable value (``best fit'') and the region around the peak
containing 68\% of all realizations is the 68\% CL uncertainty interval. 

Our analytic model for $T(r)$ allows very steep gradients. In some cases,
such profiles are formally consistent with the observed projected
temperatures because projection washes out steep gradients.  However, large
values of $dT/dr$ often lead to unphysical mass estimates, for example the
profiles with $\rho<0$ at some radius. We eliminated this problem in the
Monte-Carlo simulations by accepting only those realizations in which the
best-fit $T(r)$ leads to $\rho_{\text{tot}}>\rho_{\text{gas}}$ in the radial
range covered by the data, $r_{\text{min}}<r<r_{\text{det}}$. Finally, we
checked that the temperature profiles corresponding to the mass uncertainty
interval are all convectively stable, $d\ln T/d\ln \rho_g < 2/3$.

\begin{figure*}
\vspace*{-2.25\baselineskip}
\centerline{\includegraphics[width=1.00\linewidth]{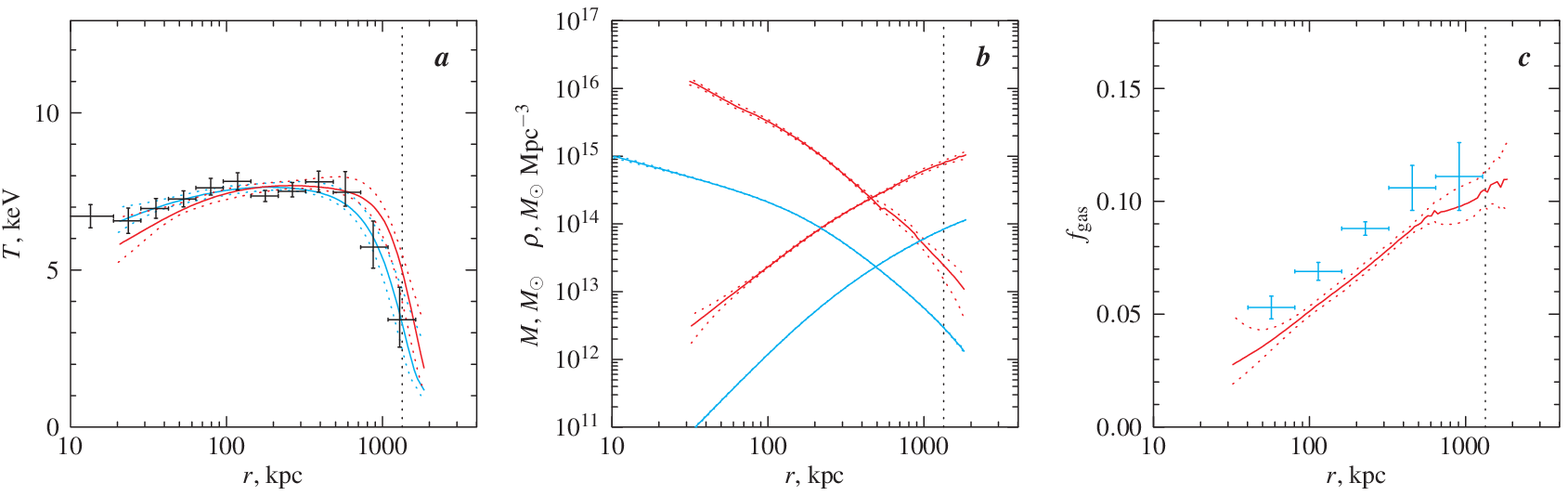}}
\vspace*{-0.5\baselineskip}
\caption{Results for A1413. See caption for Fig.\,\ref{fig:a133:res}.} 
\end{figure*}

\begin{figure*}
\vspace*{-2.5\baselineskip}
\centerline{\includegraphics[width=1.00\linewidth]{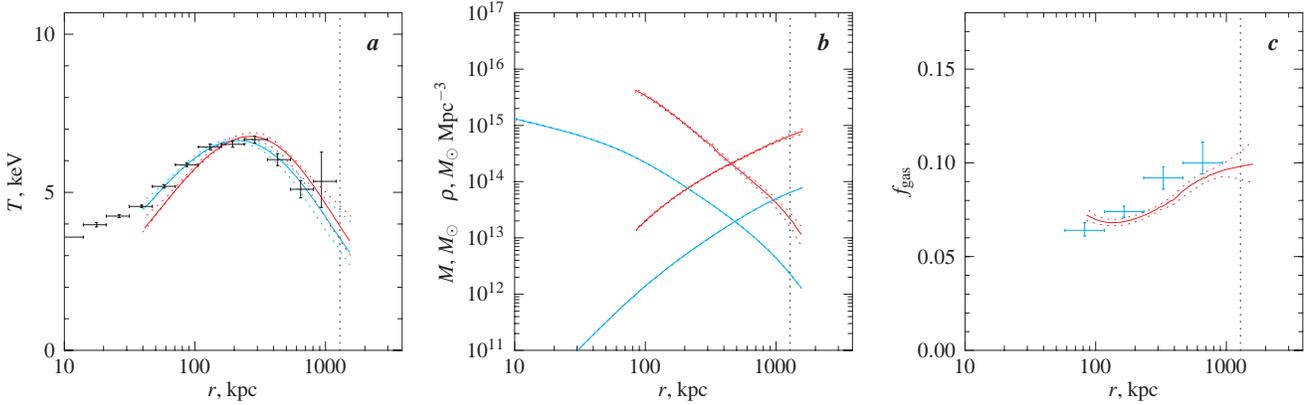}}
\vspace*{-0.5\baselineskip}
\caption{Results for A1795. See caption for Fig.\,\ref{fig:a133:res}.} 
\end{figure*}

\begin{figure*}
\vspace*{-2.5\baselineskip}
\centerline{\includegraphics[width=1.00\linewidth]{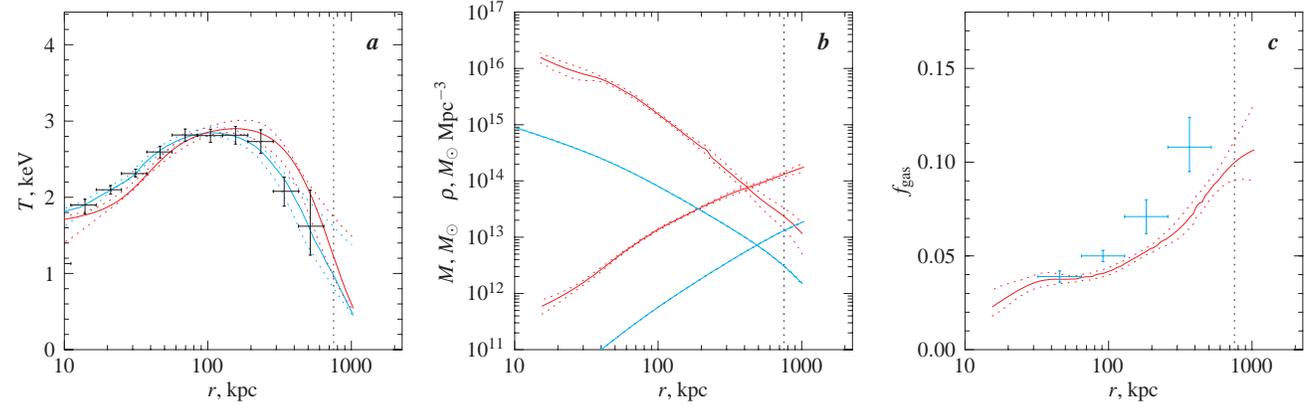}}
\vspace*{-0.5\baselineskip}
\caption{Results for A1991. See caption for Fig.\,\ref{fig:a133:res}.} 
\end{figure*}

\begin{figure*}
\vspace*{-2.25\baselineskip}
\centerline{\includegraphics[width=1.00\linewidth]{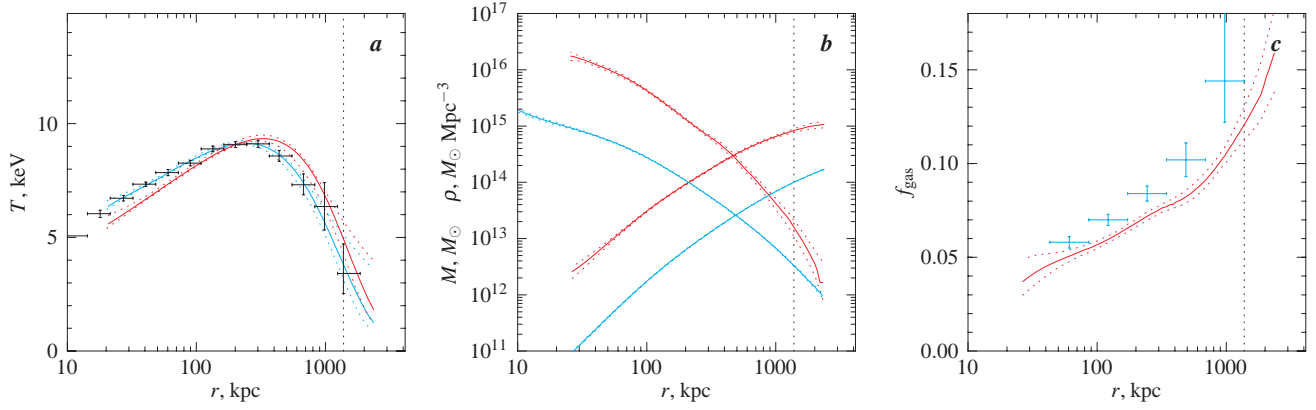}}
\vspace*{-0.5\baselineskip}
\caption{Results for A2029. See caption for Fig.\,\ref{fig:a133:res}.} 
\end{figure*}

\begin{figure*}
\vspace*{-2.5\baselineskip}
\centerline{\includegraphics[width=1.00\linewidth]{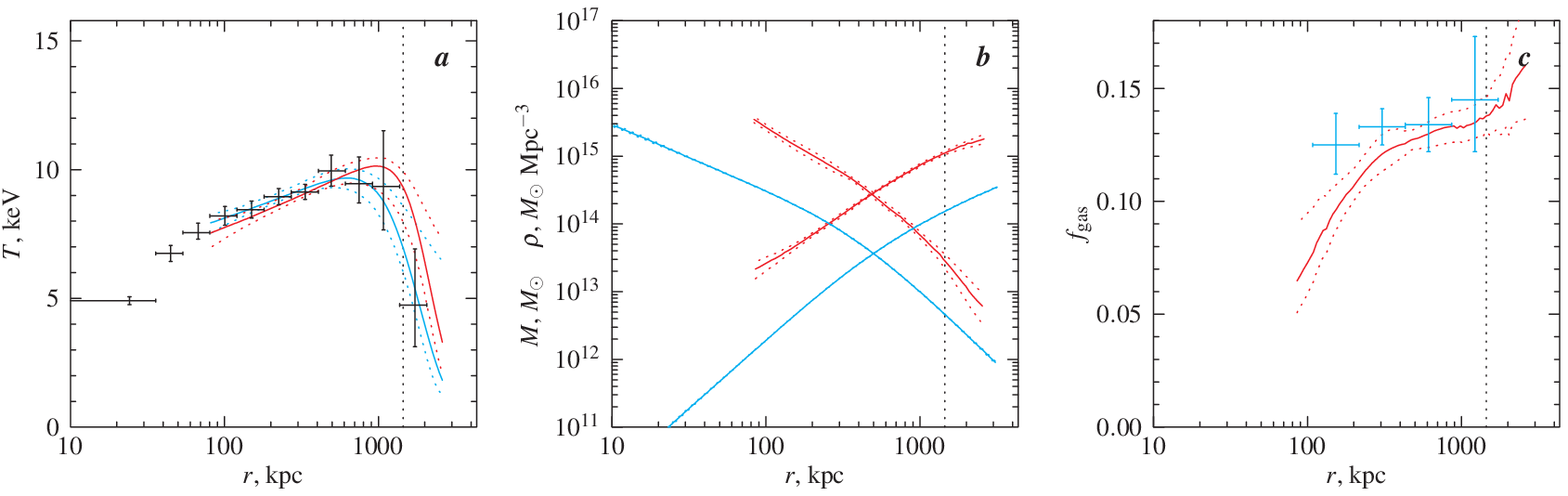}}
\vspace*{-0.5\baselineskip}
\caption{Results for A2390. See caption for Fig.\,\ref{fig:a133:res}.} 
\end{figure*}

\begin{figure*}
\vspace*{-2.5\baselineskip}
\centerline{\includegraphics[width=1.00\linewidth]{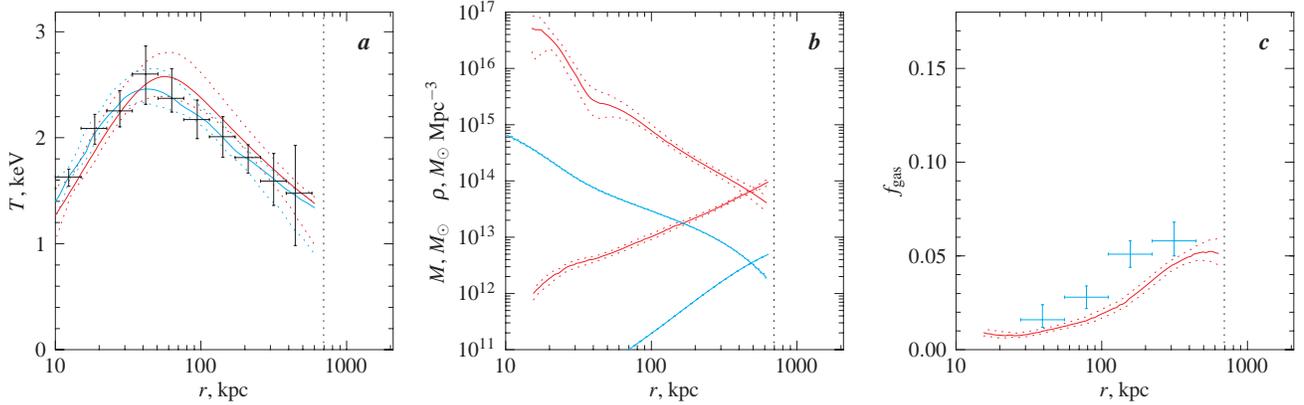}}
\vspace*{-0.5\baselineskip}
\caption{Results for RXJ\,1159+5531. See caption for Fig.\,\ref{fig:a133:res}.} 
\end{figure*}

\begin{figure*}
\vspace*{-2.25\baselineskip}
\centerline{\includegraphics[width=1.00\linewidth]{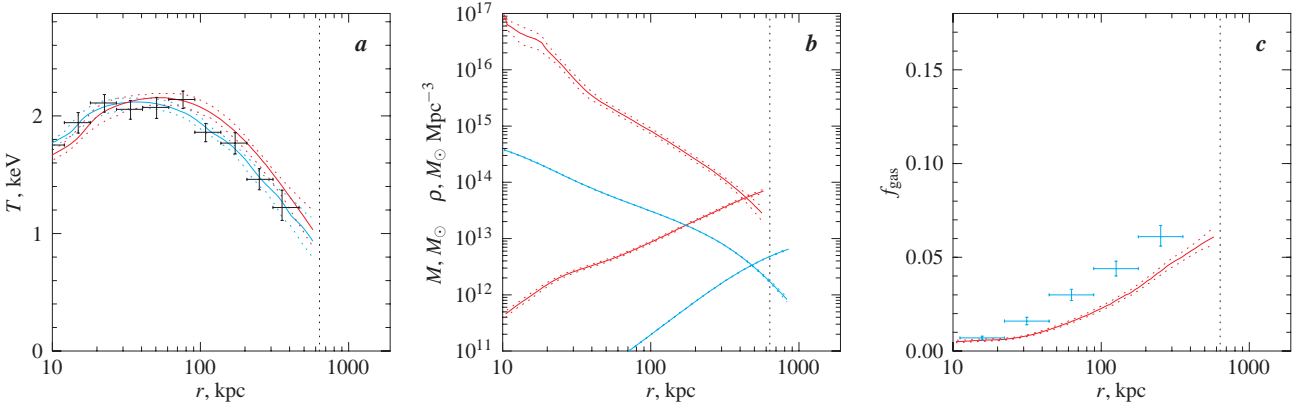}}
\vspace*{-0.5\baselineskip}
\caption{Results for MKW4. See caption for Fig.\,\ref{fig:a133:res}.} 
\end{figure*}

\begin{figure*}
\vspace*{-1.75\baselineskip}
\centerline{\includegraphics[width=1.00\linewidth]{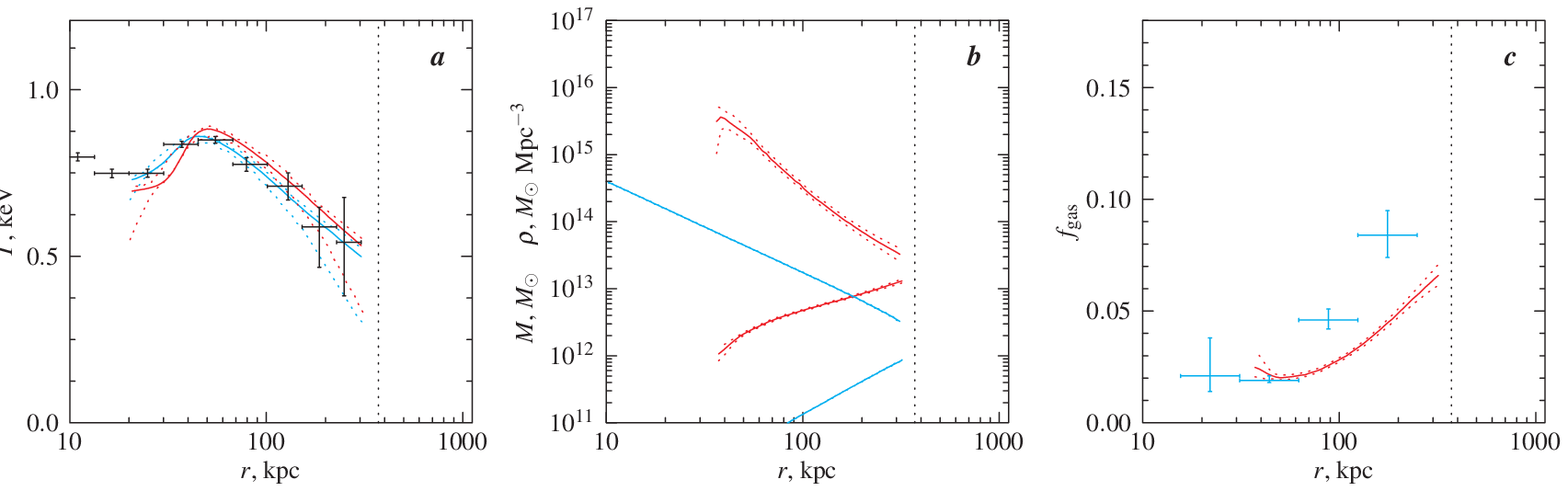}}
\vspace*{-0.5\baselineskip}
\caption{Results for USGC S152. See caption for Fig.\,\ref{fig:a133:res}.} 
\label{fig:s152:res}
\end{figure*}

\subsection{Average Temperatures}
\label{sec:mean:T}

We also computed average temperatures for each cluster using different
weightings of the 3D temperature models. The temperatures were averaged
in the radial range $70\,\text{kpc}<r<r_{500}$. The central 70~kpc were
excluded because temperatures at these radii can be strongly affected by
radiative cooling and thus not directly related to the depth of the
cluster potential well. The averages we compute are:

$\Tmg$ --- weighted with $\rho_{\text{gas}}(r)$. $\Tmg$ is needed, e.g.,
to compute the integrated SZ \citep{1972CoASP...4..173S} signal. $\Tmg$
should be more directly related to the cluster mass than X-ray
emission-weighted temperatures. 

$T_{\text{spec}}$ --- a value that would be derived from the
single-temperature fit to the total cluster spectrum, excluding the
central region.  $T_{\text{spec}}$ is obtained by integrating a
combination of $T(r)$ and $\rho_{\text{gas}}^2(r)$ as described in
\cite{astro-ph/0504098}.  $T_{\text{spec}}$ is one of the primary X-ray
observables. 

Measurement uncertainties for the average temperatures are derived via
Monte-Carlo simulations (\S\,\ref{sec:uncert:Monte-Carlo}
and~\ref{sec:mass:derivation}).

\section{Results for individual clusters}
\label{sec:results:indiv}

Best fit parameters for the gas density distribution are listed in
Table~\ref{tab:rhogas:fit}.  We do not provide measurement uncertainties
for the individual parameters because they are strongly degenerate. 
Monte-Carlo simulations show that for each cluster, the statistical
uncertainties of the gas density are below 9\% everywhere in the radial
range $5\,\text{kpc}-r_{\text{det}}$. However, extrapolations of the gas
density far beyond $r_{\text{det}}$ are unreliable. The main derived
cluster parameters are reported in Table~\ref{tab:masses}, and
temperature, mass, and gas fraction profiles for individual clusters are
presented in Fig.\,\ref{fig:a133:res}--\ref{fig:s152:res}. 

The observed temperature profile and the best-fit models are shown in
panels \emph{(a)}. The models are shown in the radial range
$r_{\text{min}}-r_{\text{det}}$. Note that our analytic function
successfully describes a wide range of shapes for the observed
temperature profiles and the corresponding projected profiles (shown by
blue lines) are always excellent fits to the data. The 3D models (red
lines) go above the data in the outer region, and below the data in the
center --- just as expected from projection of the temperature gradients
along the line of sight. 

Dotted lines show 68\% CL uncertainties for the model at each radius
derived from Monte-Carlo simulations. The uncertainty intervals are
smaller than the errorbars of the raw measurements because the model
effectively smoothes the data over 3--4 adjacent bins (and therefore
uncertainties in the neighboring bins are correlated). However, the
difference is not very large and the derived uncertainties are clearly
realistic in the sense that they include a typical range of smooth
models that could be drawn through the data.

Panels \emph{(b)} present the derived density and enclosed mass
profiles. Results for the total and gas masses are shown by red and blue
lines, respectively. $M(r)$ increases with radius and $\rho(r)$
decreases. The derived total mass profile is used to estimate the
radius, $r_{500}$ (vertical dotted lines). If $r_{500}$ is outside the
radial range covered by the X-ray data, we find it from extrapolation of
our best fit gas density and/or temperature models.  Large
extrapolations are required only for A262, RXJ\,1159+5531, and
USGC~S152. In other cases, \emph{Chandra} temperature measurements
extend almost to $r_{500}$ or beyond. 

Dotted lines show 68\% CL uncertainties for the mass and density derived
from Monte-Carlo simulations for a subset of physically meaningful
realizations (see \S\,\ref{sec:mass:derivation}). Uncertainties are
shown also for the gas mass but they are very small and almost
indistinguishable from the best fit model in these plots.

Gas fraction profiles are presented in panels \emph{(c)}. The enclosed
gas fraction is shown by red lines. The local gas fraction,
$\rho_{\text{gas}}/\rho_{\text{tot}}$, in the radial range directly
covered by the \emph{Chandra} temperature data, is shown by crosses.

Abell 2390 deserves a special note. Its deep \emph{Chandra} image
reveals large-scale cavities in the X-ray surface brightness extending
$\sim 400$~kpc from the center where a sharp break in the surface
brightness profile was reported by \cite{2001MNRAS.324..877A}. The
cavities are likely produced by bubbles of radio plasma emitted by the
central AGN, as observed in several other clusters
\citep{2005Natur.433...45M}. The ICM in the central region is not
spherically symmetric nor expected to be in hydrostatic equilibrium. 
This should result in underestimation of the total mass and
overestimation of the gas mass.  The results at small radii for A2390
should be treated with caution. There are no detectable structures
outside 500~kpc and so the results at large radii (e.q., at $r_{500}$)
should be reliable.

\section{Average Temperature Profile}
\label{sec:T:prof}

We noted in \citetalias{2004astro.ph.12306V} that the temperature
profiles for our clusters are self-similar when scaled to the same
overdensity radius, in good agreement with the earlier studies by
\citet{1998ApJ...503...77M} and \citet{2002ApJ...567..163D}. We return
to this subject here because results for individual clusters can be
compared more accurately using the reconstructed 3-dimensional
temperature profiles and the overdensity radii determined from the mass
model rather than estimated from the average temperatures as in
\citetalias{2004astro.ph.12306V}. 

\begin{figure}[b]
\vspace*{-2\baselineskip}
  \centerline{\includegraphics[width=0.95\linewidth]{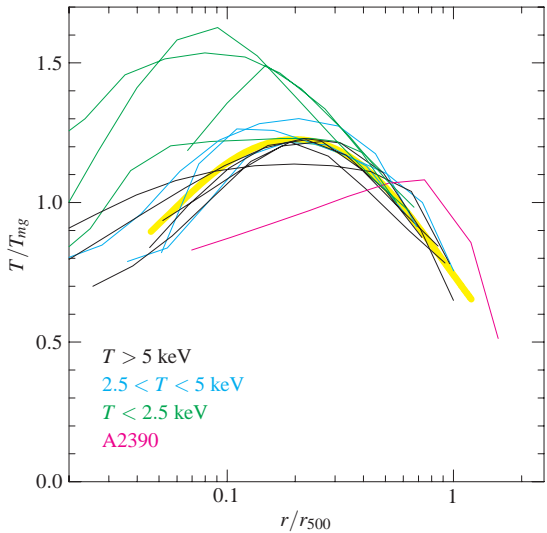}}
\vspace*{-1\baselineskip}
\caption{Scaled 3-dimensional temperature profiles. The mean temperature
  is the gas mass-weighted average. The thick yellow line shows an
  approximation to the average profile for $T>2.5$~keV clusters
  (eq.\,[\ref{eq:T:prof:average}]).} 
  \label{fig:T:prof}
\end{figure}

Figure~\ref{fig:T:prof} shows the reconstructed 3-dimensional
temperature profiles, normalized to the gas mass-weighted average
temperature, $\Tmg$, and plotted as a function of $r/r_{500}$. The
model for each cluster is plotted in the radial range directly covered
by the \emph{Chandra} spectral measurements and our 3D modeling, from
$r_{\text{min}}$ to the outer bin in
Fig.\,\ref{fig:a133:res}a--\ref{fig:s152:res}a. We normalize the profiles
to the gas mass-weighted temperature because it is less sensitive than,
e.g., $T_{\text{spec}}$ to the properties of the central regions where
the ICM temperature can be significantly affected by non-gravitational
processes. 

\def\arraystretch{1.25}
\begin{deluxetable*}{lrrrrcrccc}
\tablecaption{Masses, Average Temperatures, Concentration parameter, Gas Fractions\label{tab:masses}}
\tablehead{
\colhead{Cluster\hspace*{15mm}} &
\colhead{$r_{500}$} &
\colhead{$T_{\text{spec}}$} &
\colhead{$T_{\text{mg}}$} &
\colhead{$c_{500}$} &
\colhead{$M_{2500}$} &
\colhead{$M_{500}$} &
\colhead{$f_{\text{g, 2500}}$} &
\colhead{$f_{\text{g, 500}}$} &
\colhead{$f_{\text{g, 2500-500}}$}\\
 &\colhead{(kpc)}&\colhead{(keV)}&\colhead{(keV)}& &
 \colhead{($10^{14}\,M_\odot$)}&\colhead{($10^{14}\,M_\odot$)}
}
\startdata
A133          \dotfill & $ 998\pm37$ & $4.15\pm0.07$ & $3.68\pm0.11$ & $3.31\pm0.27$ & $1.24\pm0.08$ & $ 3.14\pm0.36$
      & $0.065\pm0.002$ & $0.084\pm0.006$ & $0.095^{+0.014}_{-0.013}$\\
A262          \dotfill & $ 663\pm19$ & $2.08\pm0.06$ & $1.92\pm0.09$ & $3.58\pm0.27$ & $0.34\pm0.05$ & \mcc{\nodata} 
      & $0.067\pm0.003$ & \mcc{\nodata}  & \mcc{\nodata} \\
A383          \dotfill & $ 956\pm33$ & $4.80\pm0.12$ & $4.36\pm0.18$ & $4.36\pm0.35$ & $1.68\pm0.15$ & $ 3.10\pm0.32$
      & $0.090\pm0.005$ & $0.122\pm0.007$ & $0.149^{+0.035}_{-0.024}$\\
A478          \dotfill & $1359\pm60$ & $7.95\pm0.14$ & $7.34\pm0.32$ & $3.75\pm0.28$ & $4.23\pm0.26$ & $ 7.83\pm1.04$
      & $0.096\pm0.004$ & $0.118\pm0.011$ & $0.130^{+0.031}_{-0.026}$\\
A907          \dotfill & $1117\pm31$ & $5.96\pm0.08$ & $5.44\pm0.13$ & $3.56\pm0.31$ & $2.30\pm0.16$ & $ 4.71\pm0.39$
      & $0.090\pm0.003$ & $0.121\pm0.005$ & $0.145^{+0.020}_{-0.016}$\\
A1413         \dotfill & $1339\pm48$ & $7.38\pm0.12$ & $6.76\pm0.20$ & $3.06\pm0.16$ & $3.08\pm0.19$ & $ 7.78\pm0.83$
      & $0.092\pm0.003$ & $0.104\pm0.007$ & $0.112^{+0.014}_{-0.014}$\\
A1795         \dotfill & $1283\pm46$ & $6.10\pm0.06$ & $5.52\pm0.11$ & $3.73\pm0.17$ & $2.66\pm0.15$ & $ 6.57\pm0.69$
      & $0.088\pm0.003$ & $0.098\pm0.007$ & $0.099^{+0.017}_{-0.013}$\\
A1991         \dotfill & $ 753\pm39$ & $2.59\pm0.06$ & $2.23\pm0.09$ & $4.66\pm0.32$ & $0.60\pm0.07$ & $ 1.28\pm0.20$
      & $0.068\pm0.004$ & $0.099\pm0.009$ & $0.103^{+0.050}_{-0.017}$\\
A2029         \dotfill & $1380\pm31$ & $8.46\pm0.09$ & $7.59\pm0.19$ & $4.13\pm0.24$ & $4.41\pm0.23$ & $ 8.29\pm0.79$
      & $0.090\pm0.003$ & $0.121\pm0.007$ & $0.144^{+0.033}_{-0.021}$\\
A2390         \dotfill & $1448\pm49$ & $8.90\pm0.17$ & $9.35\pm0.39$ & $1.65\pm0.18$ & $3.50\pm0.28$ & $10.88\pm1.05$
      & $0.127\pm0.005$ & $0.138\pm0.008$ & $0.141^{+0.014}_{-0.014}$\\
RXJ\,1159+5531\dotfill & $ 695\pm60$ & $1.80\pm0.10$ & $1.58\pm0.09$ & $1.89\pm0.35$ & $0.31\pm0.04$ & \mcc{\nodata} 
      & $0.042\pm0.002$ & \mcc{\nodata}  & \mcc{\nodata} \\
MKW4          \dotfill & $ 635\pm27$ & $1.65\pm0.05$ & $1.40\pm0.05$ & $2.69\pm0.17$ & $0.30\pm0.03$ & $ 0.74\pm0.09$
      & $0.044\pm0.002$ & $0.063\pm0.005$ & $0.073^{+0.014}_{-0.013}$\\
USGC~S152     \dotfill & \mcc{\nodata}  & $0.69\pm0.02$ & $0.59\pm0.03$ & \mcc{\nodata}  & $0.07\pm0.00$ & \mcc{\nodata} 
      & $0.043\pm0.001$ & \mcc{\nodata}  & \mcc{\nodata}
\enddata
\tablecomments{The derived quantities scale with the Hubble constant as
  $r_{500}\propto h^{-1}$, $M\propto h^{-1}$, $\fgas\propto h^{-3/2}$.
\vspace*{-1\baselineskip}
} 
\end{deluxetable*}
\def\arraystretch{1.0}

At $r\gtrsim 0.05\,r_{500}$, the scaled 3-dimensional temperature for
all but one of our $T>2.5$~keV clusters are within $\pm15\%$ from the
average profile which can be approximated with eq.\,[\ref{eq:tprof}],
\begin{equation}\label{eq:T:prof:average}
  \frac{T(r)}{\Tmg} = 1.35\;\frac{(x/0.045)^{1.9}+0.45}{(x/0.045)^{1.9}+1}\;\frac{1}{(1+(x/0.6)^2)^{0.45}},
\end{equation}
where $x=r/r_{500}$. This average model is shown in Fig.\ref{fig:T:prof}
by thick yellow line. The only deviation is A2390, in which the inner
temperature profile and overall normalization appear to be distorted by
activity of the central AGN (see above). The profiles of the
low-temperature clusters, $T<2.5$~keV (green lines in
Fig.\,\ref{fig:T:prof}), follow the average profile at $r\gtrsim
0.3\,r_{500}$ but are significantly different and show large scatter at
small radii. They show a stronger temperature increase to the center,
and the central cooler regions seem to be confined to a smaller fraction
of the virial radius than in the more massive clusters. 

The observed radial temperature variations imply that the average
cluster temperature cannot be defined uniquely. A possible difference
between different definitions (spectroscopic, emission-weighted, gas
mass-weighted etc.) should be kept in mind. Also, the aperture size used
for integration of the X-ray spectrum is important. Our average
3-dimensional profile~[\ref{eq:T:prof:average}] implies the following
approximate relation between the peak, spectroscopic average, and gas
mass-weighted temperatures (all measured in the radial range
$70~\text{kpc}-r_{500}$):
\begin{equation}
  T_{\text{peak}}:T_{\text{spec}}:\Tmg = 1.21:1.11:1. 
\end{equation}

Comparing our temperature profiles with the compilation of
\emph{XMM-Newton} results presented in \citet{2005astro.ph..2210A}, we
note a general agreement at small radii. However, the results at large
radii seem to be different --- the temperature decline is generally not
present in these \emph{XMM-Newton} clusters. Discussion of this
discrepancy is beyond the scope of this paper. The arguments for
validity of our measurements and detailed comparison with the
\citeauthor{2005astro.ph..2210A} results for several clusters in common
can be found in \citetalias{2004astro.ph.12306V}. 

\section{Total and Gas Density Profiles}
\label{sec:scaled:prof}

\begin{figure*}
  \vspace*{-4.0\baselineskip}
  \centerline{\includegraphics[width=0.75\linewidth]{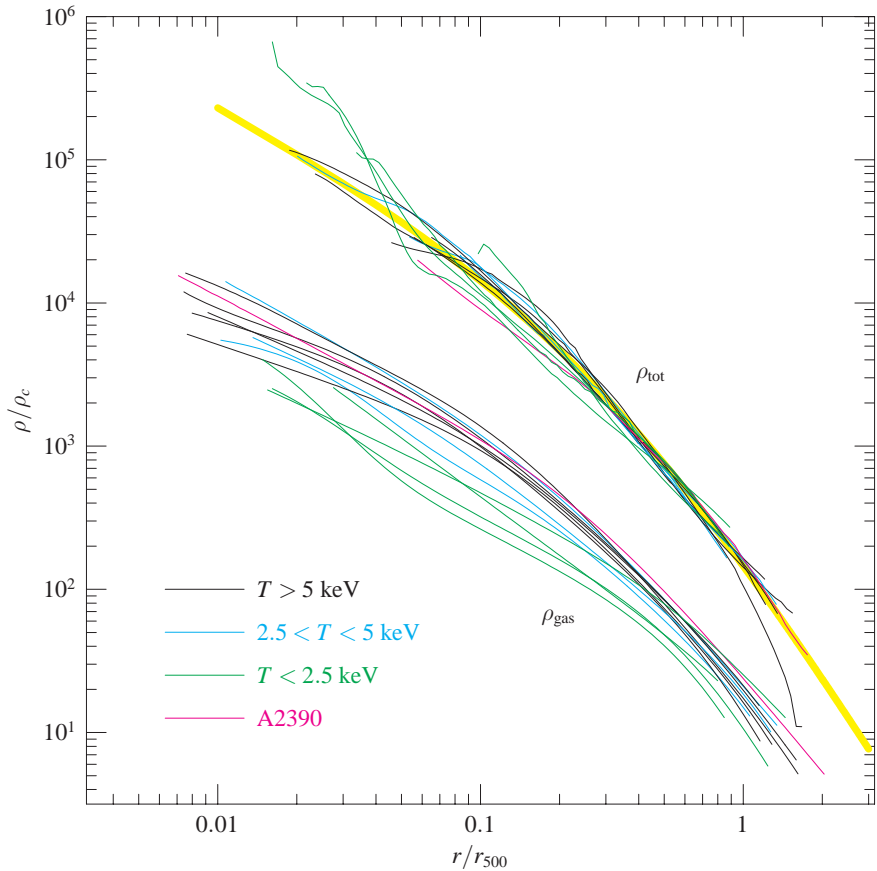}}
  \vspace*{-2.25\baselineskip}
  \caption{Scaled density profiles. Total density profiles are plotted
    within the radial range covered by the temperature profile. Gas
    density profiles are extended to $r_{\text{det}}$ (see
    Table~\ref{tab:rhogas:fit}). Thick yellow line shows the NFW model
    with $c_{500}=3$, a typical value for CDM halos in our mass range
    (\S\,\ref{sec:conc}, see Fig.\,\ref{fig:c500}).} 
  \label{fig:scaled:profs}
\end{figure*}

One of the key theoretical predictions of the hierarchical CDM models is
the universal density distribution within dark matter halos
\citep[][hereafter NFW]{navarro_etal96,navarro_etal97}. Specifically,
the shape of the radial density profiles of Cold Dark Matter halos is
characterized by a gradually changing slope $\alpha=d\log\rho/d\log r$
from $\alpha\approx -1$ in the inner regions to $\alpha\approx -3$ at
large radii \citep[NFW]{dubinski_carlberg91}. The profiles are
characterized by concentration, $c_{\Delta}$, defined as the ratio of
the halo virial radius and the scale radius, $r_s$: $c_{\Delta}\equiv
r_{\Delta}/r_s$. The scale radius is defined as the radius where the
logarithmic slope of the density profile is $\alpha=-2$.  Concentrations
of CDM halos are tightly correlated with the characteristic epoch of
object formation \citep{wechsler_etal02}.  The mean concentration is
only a weakly decreasing function of the virial mass, $c_{\Delta}\propto
M_{\Delta}^{0.1}$, \citep{navarro_etal97,bullock_etal01,eke_etal01}. 
Therefore, within a limited range of masses, the mean concentration is
approximately constant and the density profiles are approximately
self-similar. 

Such self-similarity is indeed observed in our data. In
Fig.\ref{fig:scaled:profs}, we plot the derived total density profiles,
scaled to $\rho_c(z)$, as a function of radius in units of $r_{500}$. 
The scatter of individual clusters around a mean profile is small and
consistent with that found in numerical simulations (see
\S\,\ref{sec:conc} below). The average total density profile in our
clusters agrees well with the NFW model with the concentration expected
for objects of this mass (thick yellow line in
Fig.\ref{fig:scaled:profs}). 

The $\rhotot$ profiles for 3 clusters with $T<2.5$~keV have central
steepening at $r<0.05\,r_{500}\approx 30$~kpc, which is statistically
significant. Such sharp steepening is not expected in purely CDM halos. 
We associate these components with the stellar material of the central
cD galaxies, which start to dominate the total mass at small radii. 
Indeed, our derived masses within 30~kpc for these objects, $\sim
2\times10^{12}\,M_\odot$ (see Fig.\ref{fig:a133:res}) are similar to the
stellar mass estimates in large central cluster galaxies
\citep[e.g.,][]{2004ApJ...617..879L}. 

Self-similarity is also observed in the ICM distribution at large radii. 
The lower set of profiles in Fig.\,\ref{fig:scaled:profs} shows the gas
densities, also scaled to $\rho_c(z)$. The scaled gas density near
$r_{500}$ is within $\pm15\%$ of the mean for most clusters. Some of
this scatter is caused by uncertainties in the total mass estimates ---
a typical uncertainty of 3--5\% in $r_{500}$ (Table~\ref{tab:masses})
translates into 7--12\% scatter in the scaled densities for a typical
slope of $\rhogas(r)$ near $r_{500}$. The scatter in the gas density
profiles becomes significantly larger in the inner region, which was
already noted in the previous studies
\citep[e.g.,][]{1999A&A...348..711N,1999ApJ...525...47V}. 

There is also a trend for lower-temperature clusters to have lower gas
densities and flatter profiles in the central region. This trend is
responsible for flat gas density slopes derived for galaxy groups and
low-mass clusters in the previous analyses using the $\beta$-model
approximations
\citep[e.g.,][]{2000MNRAS.315..356H,2001A&A...368..749F,2003MNRAS.340..989S}. 
However, we observe that at large radii, the gas density in cold
clusters approaches the average profile defined by $T>5$~keV clusters. 
Also, the gas density profiles often steepen at $r\simeq 0.7\,r_{500}$
so that the slopes near $r_{500}$ are similar for all clusters. Further
discussion of this issue and its impact on the hydrostatic mass
estimates is presented in Appendix~\ref{sec:MT:comparisons}. 

Below, we compare the concentration parameters for our \rhotot{} profiles
with those expected for CDM halos. Comparison of the gas density
distributions with the results of numerical simulations will be
presented in a future paper. 

\subsection{Concentration parameters}
\label{sec:conc}

\begin{figure}[b]
  \vspace*{-2.5\baselineskip}
  \centerline{\includegraphics[width=0.999\linewidth]{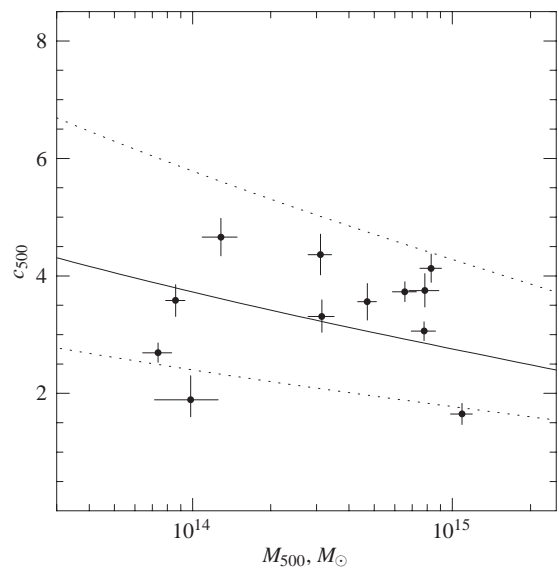}}
  \vspace*{-1.0\baselineskip}
  \caption{Concentration parameters of the NFW model,
    $c_{500}=r_{500}/r_s$, as a function of cluster mass. Solid points
    with error bars are our measurements. Solid line shows the average
    concentration of CDM halos from simulations by \citet{dolag_etal04}. 
    Dotted lines show 2-$\sigma$ scatter of log-normal distribution of
    concentrations at a fixed mass, found in simulations.} 
  \label{fig:c500}
\end{figure}

As discussed above, the $\Lambda$CDM paradigm makes a firm theoretical
prediction for concentrations of the dark matter halos. It is
interesting to compare our measurements with these predictions. We
define the concentration as $c_{500}\equiv r_{500}/r_s$ because our mass
measurements typically extend to $\sim r_{500}$. The scale radius $r_s$
was determined by fitting the NFW model to values of the total density
at six radii equally log-spaced in the range
$0.05\,r_{500}-r_{\text{det}}$. 
% with weights given by the uncertainty of
% $\rho_{\text{tot}}$ at these radii. 
The range $r<0.05\,r_{500}$ is excluded because at these radii there are
separate mass components associated with the stellar material in cD
galaxies (see above); such components are found in a similar radial
range in numerical simulations, which include cooling and star formation
\citep[e.g.,][]{2004ApJ...616...16G}. The uncertainties for $c_{500}$
are derived from the Monte-Carlo simulations described in
\S\,\ref{sec:prof:modeling}. The resulting $c_{500}$ determinations are
reported in Table~\ref{tab:masses} and plotted as a function of measured
$M_{500}$ in Fig.\ref{fig:c500}. 

\begin{figure*}
\vspace*{-2.5\baselineskip}
\centerline{\mbox{}\hfill%
  \includegraphics[width=0.478\linewidth]{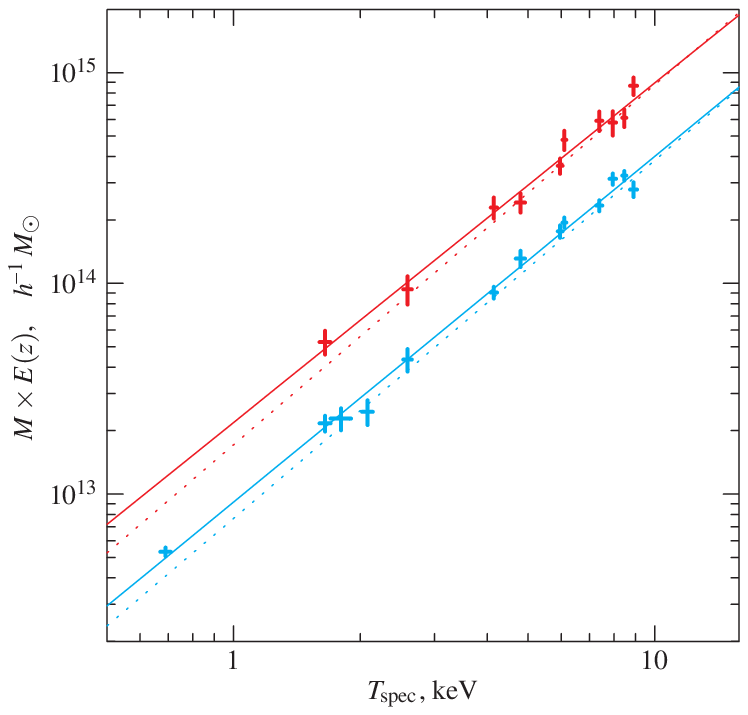}\hfill%
  \includegraphics[width=0.478\linewidth]{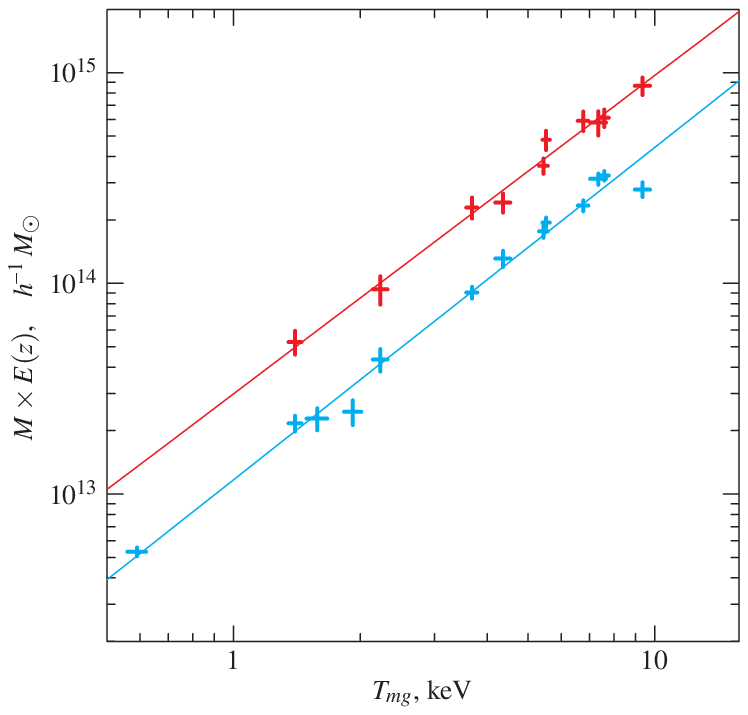}\hfill\mbox{}%
}
\vspace*{-1.0\baselineskip}
\caption{Total mass within $r_{500}$ (red) and $r_{2500}$ (blue), as a
  function of X-ray spectroscopic temperature, \Tspec, and gas
  mass-weighted temperature, $\Tmg$. Solid lines show the best fit power
  laws with parameters listed in Table~\ref{tab:mtfit}.  Dotted lines in
  the \Tspec{} plot show the best fit relations from
  \citet{2005astro.ph..2210A}. Note that $M_{500}$ measurements are not
  plotted for clusters which required large extrapolations of the
  temperature profiles (A262, RXJ\,1159+5531, USGC~S152).} 
\label{fig:MT:Tspec}
\label{fig:MT:Tmgas}
\end{figure*}

The measurements are compared with the expected $c(M)$ relation
suggested by the cluster simulations of \citet[][see their eq.~13 and
Table~2]{dolag_etal04} in the ``concordance'' $\Lambda$CDM cosmology,
$\Omega_0=1-\Omega_{\Lambda}=0.3$, $\sigma_8=0.9$ (solid line in
Fig.\,\ref{fig:c500}). Note that we converted the concentrations and
masses from the definitions used by \citeauthor{dolag_etal04} to
$c_{500}$ and $M_{500}$ used in our analysis
\citep[e.g.,][]{2003ApJ...584..702H}. We also show (dotted lines) the
2-$\sigma$ scatter of concentrations typically found in numerical
simulations, $\sigma_{\ln c}=0.22$
\citep{jing00,bullock_etal01,wechsler_etal02,tasitsiomi_etal04,dolag_etal04}. 

Clearly, both the typical values and scatter of concentrations
determined for our clusters are in general agreement with the simulation
results.  It can be argued that for massive clusters, most of our
measurements are slightly higher than the theoretical average. If this
effect is real, it can be caused by several factors. First, our sample
contains only a highly relaxed sub-population of galaxy clusters which
are expected to sample the high tail of the concentration distribution
\citep{wechsler_etal02}.  Second, radiative cooling of baryons and the
associated galaxy formation are expected to modify the density profiles
of parent CDM halos. These processes are expected to steepen the mass
distribution in the inner regions of clusters, due to buildup of the
central cluster galaxy \citep{2004ApJ...616...16G} and make the mass
distribution considerably more spherical on scales as large as $r_{500}$
\citep{kazantzidis_etal04}.  The amplitude of this effect, determined
from a sample of simulated clusters described in
\citet*{kravtsov_etal05} with two additional massive Coma-size objects,
is $\Delta c_{500} = 0.5-1$ (Kravtsov et al., in preparation). This is
similar to the possible systematic enhancement of concentration we observe
for $M>2\times10^{14}\,M_\odot$ clusters. 

\begin{deluxetable*}{lccccccccc}
\tablecaption{Power Law Fit to Mass-Temperature Relation\label{tab:mtfit}}
\tablehead{
 &
\multicolumn{3}{c}{$T_{\text{spec}}$} &&
\multicolumn{3}{c}{$T_{\text{mg}}$} \\[2pt]
\cline{2-4}
\cline{6-9}\\[-6pt]
\colhead{Overdensity\hspace*{15mm}}  &\colhead{$M_5$,~~$h^{-1}M_\odot$}&\colhead{$r_5$,~~$h^{-1}$~Mpc}&\colhead{$\alpha$}&&\colhead{$M_5$,~~$h^{-1}M_\odot$}&\colhead{$r_5$,~~$h^{-1}$~Mpc}&\colhead{$\alpha$}}
\startdata
$\Delta=500$~\dotfill & $(2.93\pm0.16)\times10^{14}$ & $0.796\pm0.015$ & $1.61\pm0.11$ && $(3.41\pm0.18)\times10^{14}$ & $0.837\pm0.015$ & $1.51\pm0.10$\\
$\Delta=2500$~\dotfill & $(1.28\pm0.05)\times10^{14}$ & $0.354\pm0.005$ & $1.64\pm0.06$ && $(1.48\pm0.07)\times10^{14}$ & $0.371\pm0.006$ & $1.58\pm0.07$
\enddata
\tablecomments{We use a power law fit of the form
  $h(z)\,M=M_5\times(T/5\,\text{keV})^{\alpha}$, where $M=M_{500}$ or
  $M_{2500}$ and temperatures are either X-ray spectral or gas
  mass-weighted averages (see \S\,\ref{sec:mean:T}). Scaling of the
  corresponding overdensity radii with temperature is of the
  form~[\ref{eq:RT}].
\vspace*{-\baselineskip}
} 
\end{deluxetable*}

% \***{Check $\rho \times r^2$ and see if there is a maximum. --- If we
%   see it, we measure concentration more reliable than the others (who
%   rely on the central change in slope from -1 to -2).} 
% 

The main conclusion from this analysis is that there is good overall
agreement between theoretical expectations and the measured
concentration parameters in our cluster sample. A similar conclusion was
reached by \citet{2005A&A...435....1P} from the \emph{XMM-Newton}
analysis of 10~clusters. 

\section{Mass-Temperature Relation}
\label{sec:MT}

A tight relation between the cluster temperature and total mass is
expected on theoretical grounds, which is indeed observed at least for
the hydrostatic mass estimates
\citep{2000ApJ...532..694N,1999ApJ...520...78H,2001A&A...368..749F,2001ApJ...553...78X,2003MNRAS.340..989S}. 
Recent determinations of the $M-T$ relation from \emph{Chandra} and
\emph{XMM-Newton} observations were presented in
\citet{2001MNRAS.328L..37A} and \citet{2005astro.ph..2210A},
respectively (\citeauthor{2001MNRAS.328L..37A} derived masses for the
critical overdensity 2500 and \citeauthor{2005astro.ph..2210A} also
presented measurements at larger radii, including extrapolation to
$r_{200}$). We can significantly improve over these previous
measurements because our temperature profiles extend to large radii and
therefore we do not use simplifying assumptions such as that $T(r)$ is
constant or polytropic.  Also, we have high-quality X-ray surface
brightness measurements for all clusters at radii well beyond $r_{500}$,
and do not use the common $\beta$-model approximation to derive the gas
densities. 

We measure total masses for two often-used critical overdensity levels,
$\Delta=2500$, and $\Delta=500$. The $\Delta=500$ level is particularly
useful because it approximately delineates the inner cluster region
where the bulk ICM velocities are small and therefore the hydrostatic
mass estimates are meaningful \citep*{1996ApJ...469..494E}. The
corresponding radius, $r_{500}$, is 0.5--0.67 of the virial radius,
depending on $\Omega_m$ \citep{eke_etal96,bryan_norman98}. The
$\Delta=2500$ overdensity level encompasses the bright central region
where X-ray temperature profile measurements are feasible with
\emph{Chandra} even in high-redshift clusters
\citep{2004MNRAS.353..457A}. 

We do not consider the masses for lower overdensities because this would
require extrapolation far beyond the radius covered by the
\emph{Chandra} data, and because the ICM is not expected to be fully in
hydrostatic equilibrium at large radii. Mass measurements are sometimes
extrapolated from the inner region to, e.g., $r_{200}$ assuming the NFW
model for the matter density profile. Such extrapolations are highly
model-dependent and lead to underestimated measurement uncertainties for
$M_{200}$. It is more appropriate to scale the theoretical models to
$r_{500}$, where direct measurements are now available. Scaling of mass
of CDM halos to any overdensity level is straightforward
\citep{2003ApJ...584..702H}. 

The measured values of $M_{2500}$ and $M_{500}$ are reported in
Table~\ref{tab:masses}. We do not compute $M_{500}$ for three clusters
for which the temperature profile measurements have to be extrapolated
too far (A262, RXJ\,1159+5531, and USGC~S152). For A2390, $M_{2500}$
should be treated with caution because $r_{2500}$ is near the boundary
of the central non-hydrostatic region (see \S\,\ref{sec:results:indiv}). 

In Fig.\ref{fig:MT:Tspec}, we plot the measured masses as a function of
the X-ray spectroscopic and gas mass-weighted temperatures. The
self-similar evolution of the normalization of the $M-T$ relation is
expected to follow
\begin{equation}
  M_\Delta/T^{3/2} \propto E(z)^{-1}, \quad E(z)=H(z)/H_0,
\end{equation}
\citep[e.g.,][]{bryan_norman98}. Even though the effect of evolution is
small within our redshift interval, we applied the corresponding
corrections by multiplying the measured masses by $E(z)$ computed for
our preferred cosmology, thus ``adjusting'' all measurements to $z=0$.

We fit the observed $M-T$ relation with a power law,
\begin{equation}
  M = M_5 \, (T/5\,\text{keV})^\alpha. 
\end{equation}
The relation is normalized to $T=5$~keV because this is approximately
the median temperature for our sample and therefore the estimates for
$M_5$ and $\alpha$ should be nearly independent. The fit is performed
using the bisector modification of the \citet[and references
therein]{1996ApJ...470..706A} linear regression algorithm that allows
for intrinsic scatter and nonuniform measurement errors in both
variables. The uncertainties were evaluated by bootstrap resampling
\citep[e.g.,][]{numerical_recipes}, while simultaneously adding random
measurement errors to $M$ and $T$. 

% Slopes:
%        Horner:    1.5 for ``spat. resolved'', 1.9 for isothermal
%        Jukka:     1.8
%        Fenka:     1.8
%        Xu:        1.6
%        Allen:     does not constrain
%        Sanderson: 1.85-1.9
%        Monique:   1.7

The best fit slopes and normalizations for different variants of the
$M-T$ relations are reported in Table~\ref{tab:mtfit}. We find flatter
slopes ($\alpha\simeq1.5-1.6$) than many previous studies (typically,
$\alpha=1.7-1.8$ when low-temperature clusters were included). Some of
the difference can be traced to slightly different definitions of the
cluster temperature in individual studies and different procedures for
scaling the measurements to $z=0$. However, the major effects are
accurate measurements of the temperature gradient at large radii and
correct modeling of the steepening in the gas density profiles at large
radii. The comparison with other works discussed in detail in
Appendix~\ref{sec:MT:comparisons}. 

The $M-T$ relation implies the following scaling of the overdensity
radii with temperature
\begin{equation}\label{eq:RT}
  r_\Delta\,h\,E(z) = r_5\, (T/5\,\text{keV})^{\alpha/3},
\end{equation}
with coefficients $r_5$ provided in Table~\ref{tab:mtfit}. 

As seen in Fig.\ref{fig:MT:Tspec}, the scatter of the individual $M$ and
$T$ measurements around the best fit power law approximations is very
small. Note that in our case, unlike many previous studies, this is not
just a trivial consequence of the approximate similarity of the cluster
X-ray surface brightness profiles \citep{1999A&A...348..711N} because we
do not use overly-constrained models for $T(r)$. To characterize the
scatter, we compute the rms deviations in mass and subtract the expected
contribution from the measurement errors,
\begin{equation}
  (\delta M/M)^2 = \frac{1}{N-2} \sum \frac{(M_i - M_5 (T_i/5)^{\alpha})^2 - \Delta M_i^2}{M_i^2},
\end{equation}
where $\Delta M_i$ are measurement errors and $\delta M/M$ is the
estimated scatter in the relation (the $1/(N-2)$ factor accounts for 2
degrees of freedom in the power law fit). We find that the observed
scatter is consistent with zero, and the 90\% upper limits for intrinsic
scatter are $\delta M/M \simeq 0.15$ for both $M_{500}$ and $M_{2500}$,
and for both definitions of the mean temperature, \Tspec{} and \Tmg. 
Note the scatter is likely to be larger for the whole cluster population
including non-relaxed objects. 

Our normalizations of the $M-T$ relation are higher than most of the
previous X-ray determinations based on \emph{ASCA} and \emph{ROSAT}
analyses (see Appendix~\ref{sec:MT:comparisons} for detailed
discussion). There is, however, a very good agreement with the
\emph{XMM-Newton} measurements by \citet[their results are shown by
dotted lines in Fig.\ref{fig:MT:Tspec}]{2005astro.ph..2210A}, although
this appears to be a result of a chance cancellation of systematic
differences between our analyses
(Appendix~\ref{sec:comparison:arnaud.etal}). 

Our normalization of the $M-T$ relation is also in good agreement with
those derived from recent high-resolution numerical simulations
\citep[e.g.,][]{2004MNRAS.348.1078B} that attempt to model
non-gravitational processes in the ICM (radiative cooling, star
formation, feedback from SN). Detailed comparison of our measurements
with the results of numerical simulations will be presented elsewhere.

The primary goal of observationally calibrating the $M-T$ relation is
for use in fitting cosmological models to the cluster temperature
function. Given the good agreement of our results with some other recent
measurements \citep{2005astro.ph..2210A} and results of realistic
cluster numerical simulations, it is tempting to conclude that
observational determinations of the $M-T$ normalization have finally
converged to the true value. We, however, caution against direct
application of our normalizations to fitting the published cluster
temperature functions
\citep{1998ApJ...504...27M,2000ApJ...534..565H,2002A&A...383..773I} for
several reasons. First, the definitions of the mean cluster temperature
in these papers is not identical to ours.  Second, our determination of
the $M-T$ relation uses only the most relaxed clusters at the present
epoch and it can be significantly different for mergers
\citep{2002ApJ...577..579R}. Even for relaxed clusters in the present
sample, our analysis neglects potential deviations from hydrostatic
equilibrium (e.g., caused by ICM turbulence) or deviations from
spherical symmetry. A detailed study of these effects will be presented
in Nagai et al.{} (in preparation); they can lead to $\sim 10\%$
underestimation of the total masses in our analysis
\citep[e.g.,][]{1996ApJ...469..494E,2004MNRAS.351..237R,2004MNRAS.355.1091K,faltenbacher_etal05}. 

\begin{figure}[b]
  \vspace*{-2.5\baselineskip}
  \centerline{\includegraphics[width=0.95\linewidth]{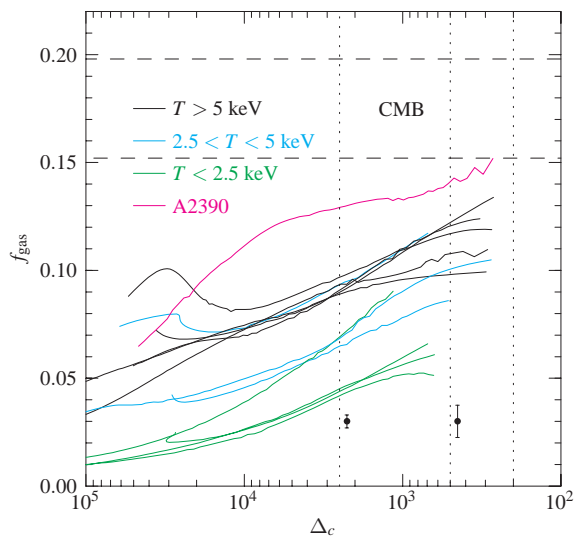}}
  \vspace*{-1.0\baselineskip}
  \caption{Enclosed gas fraction as a function of overdensity, defined
    with respect to the critical density at the cluster redshift. Dashed
    lines show the range $\Omega_b/\Omega_m=0.175\pm0.023$ constrained
    by CMB observations. Points with error bars show typical measurement
    uncertainties at two radii.} 
  \label{fig:fgas:delta}
\end{figure}

\section{Gas fractions}
\label{sec:fgas}

Finally, we present determinations of the ICM mass fractions in our
clusters. Such measurements, especially within a large fraction of the
cluster virial radius, are of great cosmological significance because
they provide an independent test for $\Omega_M$ assuming that the baryon
fraction in clusters is close to the cosmic mean \citep{1993Natur.366..429W}. 

The derived ICM mass fractions, \fgas, as a function of radius are shown
for individual clusters in
Fig.\,\ref{fig:a133:res}c--\ref{fig:s152:res}c. We present results for
both enclosed gas fractions (continuous lines) and local fractions
within several spherical shells (crosses). In almost all clusters there
is a significant increase of \fgas{} with radius, in qualitative
agreement with a number of previous measurements that used
spatially-resolved temperatures
\citep[e.g.,][]{1997ApJ...491..467M,1999ApJ...527..545M,2002A&A...394..375P,2004MNRAS.353..457A}. 

Comparison of observed gas fractions for individual clusters as a
function of overdensity (Fig.\ref{fig:fgas:delta}) gives further
insights. At $\Delta<10^{4}$, the gas fraction increases with radius as
a power law of overdensity. Except for maybe in three clusters, no
flattening of $\fgas(r)$ is observed, at least within $r_{500}$. The
average ratio of \fgas{} within $r_{2500}$ and $r_{500}$ is
$f_{2500}/f_{500}=0.84$ in massive clusters.

\begin{figure}
  \vspace*{-2.5\baselineskip}
  \centerline{\includegraphics[width=0.95\linewidth]{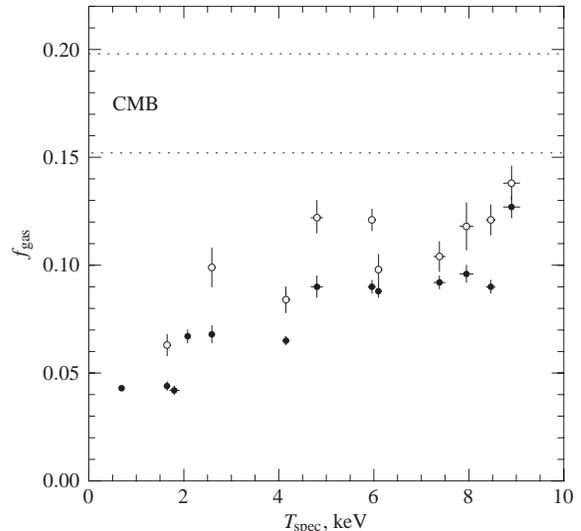}}
  \vspace*{-1\baselineskip}
  \caption{Observed gas fractions within $r_{2500}$ and $r_{500}$ (solid
    and open circles, respectively). We do not compute $f_{500}$ for
    A262, RXJ\,1159+5531, and USGC~S152 because this requires large
    extrapolations of the mass model. The $f_{2500}$ value for A2390
    ($T=8.9$~keV), can be biased (\S\,\ref{sec:results:indiv}). }
  \label{fig:fgas:T}
\end{figure}

There is also a significant trend of increasing \fgas{} with cluster
mass. The effect is the strongest at small radii, but even within
$r_{500}$, $\fgas(M)$ is not constant. In Fig.\ref{fig:fgas:T}, we plot
the enclosed gas fractions within $r_{2500}$ and $r_{500}$ as a function
of cluster temperature. The gas fraction at $r_{2500}$ increases
approximately linearly with temperature from $f_{2500}\simeq 0.04$ for
three $T<2$~keV clusters to $f_{2500}\simeq 0.10-0.11$ for the most
massive clusters. Another possibility is that there is a flattening at
$f_{2500}\simeq 0.09$ at $T>5$~keV and that the measurement for the
highest-temperature cluster, A2390, is strongly biased (see
\S\,\ref{sec:results:indiv}). We cannot distinguish these possibilities
because of significant object-to-object scatter and the small sample
size. 

The global baryon fraction in the Universe is constrained by CMB
observations to be $\Omega_b/\Omega_m = 0.175\pm 0.023$
\citep{2004Sci...306..836R,2003ApJS..148..175S}. Therefore, the observed
gas fraction within $r_{2500}$, even in the most massive of our
clusters, is significantly lower, by a factor of $\sim 0.6$, than the
cosmic mean. A similar level of baryonic deficit was previously noted by
\citet{2003MNRAS.344L..13E} and its systematic variation with mass was
obtained indirectly by \citet{1999MNRAS.305..631A} and
\citet{1999ApJ...517..627M}. This deficit, and cluster-to-cluster
variations of $f_{2500}$ can, at least in part, be explained by
conversion of ICM into stars. Contribution of the stars to the total
baryon budget should be most important in the cluster centers because of
the presence of cD galaxies.  Indeed, the largest cD galaxies have
K-band luminosities $(1-2)\times10^{12}\,L_\odot$
\citep{2004ApJ...617..879L}. Assuming a stellar mass-to-light ratio in
the K-band of $\sim 1$ \citep{2003ApJS..149..289B}, we can estimate that
just the cD, not counting other galaxies, can contribute from $0.07$ to
$0.01$ to the \emph{baryon} fraction within $r_{2500}$, depending on the
cluster mass.  Therefore, the stellar mass in the cluster centers is
significant and it should be determined individually in each cluster. 
Another important process that can affect \fgas{} in the cluster centers
--- and otherwise break the self-similarity of the gas properties --- is
energy output from central AGNs \citep{2005ApJ...625L...9N}. 

The stellar contribution to the total baryon mass and relative
energetics of non-gravitational processes should be less important
within larger radii. Indeed, we observe that \fgas{} increases between
$r_{2500}$ and $r_{500}$, by factors of 1.2--1.4. Also, the trend of
$f_{500}$ with temperature is weaker than that for $f_{2500}$. It is
consistent with both a linear increase of \fgas{} with $T$ and
flattening at $T>5$~keV. 

\begin{figure}
  \vspace*{-2.5\baselineskip}
  \centerline{\includegraphics[width=0.95\linewidth]{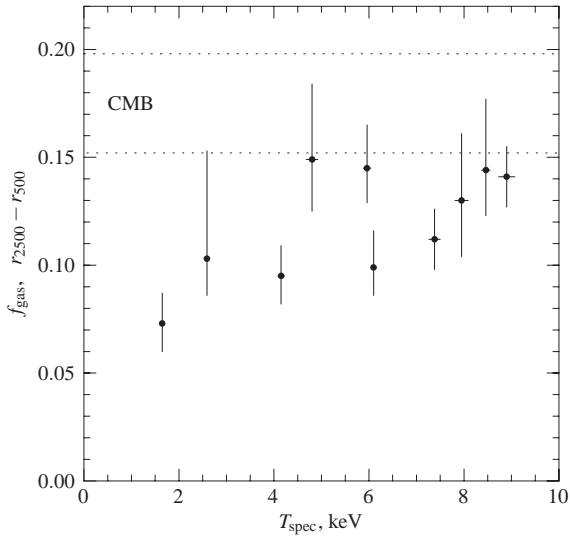}}
  \vspace*{-1\baselineskip}
  \caption{Observed gas fractions within the shell $r_{2500}$--$r_{500}$. 
    We do not compute \fgas{} for A262, RXJ\,1159+5531, and USGC~S152
    because this requires large extrapolations of the mass model.} 
  \label{fig:fgas:shell:T}
\end{figure}

Note, however, that typically, 40--50\% of the total mass within
$r_{500}$ is contained within the region $r<r_{2500}$ affected by
stellar contribution and non-gravitational heating by AGNs. One can hope
that a greater uniformity is observed if this central region is removed
entirely from the calculation of \fgas. In Fig.\ref{fig:fgas:shell:T},
we plot gas fractions measured in the $r_{2500}$--$r_{500}$ shell. The
gas fractions in the shell increase still further relative to $f_{500}$. 
In fact, they become consistent (albeit within larger uncertainties)
with an average of $\fgas\simeq 0.13$ for all clusters, except for the
lowest-temperature object (MKW~4). A reasonable correction for the
stellar contribution, 10--15\% of the ICM mass
\citep*[e.g.,][]{2003ApJ...591..749L,2003MNRAS.345.1241S,2004ApJ...601..610V},
should then bring the gas fractions within the shell
$r_{2500}$--$r_{500}$ into agreement with the Universal value determined
from CMB studies.

To summarize, the observed gas fraction shows smaller variations between
individual objects, and is closer to the Universal value, when
determined at larger radii. This is in line with the general tendency
for our clusters to become more self-similar at large radii, as
manifested by the 3-dimensional temperature and density profiles, and
the $M-T$ relation. The regularity of \fgas{} determined in the shell
$r_{2500}$--$r_{500}$ gives hope that these measurements can be used to
determine $\Omega_m$ using the classical baryon fraction test
\citep{1993Natur.366..429W}. We defer application of this test to a
future work because this involves small but important corrections, such
as stellar contribution, baryon depletion, accuracy of hydrostatic
assumption etc., which are beyond the scope of this paper. 

We finally mention that gas fractions within $r_{2500}$ derived from
\emph{Chandra} observations of relaxed clusters were recently used for
cosmological constraints in a series of papers by \citet[and references
therein]{2004MNRAS.353..457A}.  There are significant differences in the
$f_{2500}$ derived in our analysis and those reported by
\citeauthor{2004MNRAS.353..457A}, as outlined in
Appendix~\ref{sec:comp:allen}. 

\section{Summary and future work}

We presented total mass and gas profiles of low-redshift, relaxed
clusters using the best available observations with \emph{Chandra}. The
cluster sample (13 objects) spans a range of temperatures of 0.7--9~keV
and masses $M_{500}=(0.5-10)\times10^{14}\,M_\odot$. The total masses
are derived assuming hydrostatic equilibrium of the ICM in the cluster
potential.  Our modeling method makes few additional assumptions and
allows us to reconstruct the cluster properties relatively
model-independently, thus avoiding biases and producing realistic
uncertainties on the derived quantities.  The main results are
summarized below.

1. The shape of the total density profiles in our clusters and their
scaling with mass are in good agreement with predictions of $\Lambda$CDM
model. They approximately follow the universal density profile with the
concentration expected for CDM-dominated halos in this cosmology. The
gas density and temperature profiles at large radii are also nearly
self-similar, although in the inner region (within $\approx
0.1~r_{500}$) there is significant scatter and systematic trend with
cluster mass. 

2. Correspondingly, we find that the slope of the mass-temperature
relation is in good agreement with the simple self-similar behavior,
$M_{500}\propto T^\alpha$ where $\alpha=(1.5-1.6)\pm0.1$, if the average
temperature is measured at radii not affected by the central cool core. 
We derive an accurate normalization of the $M-T$ relation for relaxed
clusters. Our normalization is $\approx 30\%$ higher than most previous
X-ray determinations. 

3. The gas density profiles are generally not described by a
$\beta$-model or its common modifications, even at large radii. The
profiles steepen continuously and this behavior can be approximated as a
smooth break in the power law index around $(0.5-0.7)\,r_{500}$.  Near
$r_{500}$, the effective slope of the gas density profile,
$-\frac{1}{3}\,d\log \rho_g/d\log r \simeq 0.78$, with no detectable
dependence on cluster mass. 

This behavior of the gas density at large radii is missed by
$\beta$-model fits because they are primarily sensitive to the data in
the bright inner region. Insufficiently accurate modeling of the gas
distribution at large radii, in addition to using polytropic
approximations to $T(r)$, explains the lower normalizations of the $M-T$
relation derived in previous studies. 

4. We present accurate measurements of the gas mass fraction as a
function of radius.  We observe strong systematic variations of \fgas{}
both with radius and with cluster mass.  The gas fractions within
$r_{2500}$ are significantly lower than the Universal baryon fraction
suggested by the CMB observations. However, the trends become weaker and
the absolute values of \fgas{} are closer to the Universal value at
$r>r_{2500}$. 

In future work, we will use these accurate measurements of the gas
fractions to constrain $\Omega_m$. This requires the inclusion of small
but important effects, such as stellar mass, baryon depletion, and
correction for biases in the mass measurements, which are beyond of the
scope of this paper. We will also present a detailed comparison of the
\emph{Chandra} results on the cluster mass, temperature, and \fgas{}
profiles with high-resolution cosmological simulations. 

\acknowledgements

We thank E.~Pointecouteau, M.~Arnaud, and S.~Allen for sharing details
of their analyses%
. 
% and acknowledge useful discussions with .... 
This work was supported by NASA grant NAG5-9217 and contract NAS8-39073,
and the Smithsonian Institution.  AVK was supported by NSF grants
AST-0206216 and AST-0239759, by NASA grant NAG5-13274, and by the Kavli
Institute for Cosmological Physics at the University of Chicago.

\bibliographystyle{apj}
\bibliography{mprof}

\begin{thebibliography}{89}
\expandafter\ifx\csname natexlab\endcsname\relax\def\natexlab#1{#1}\fi

\bibitem[{{Akritas} \& {Bershady}(1996)}]{1996ApJ...470..706A}
{Akritas}, M.~G. \& {Bershady}, M.~A. 1996, \apj, 470, 706

\bibitem[{{Allen} et~al.(2001{\natexlab{a}}){Allen}, {Ettori} \&
  {Fabian}}]{2001MNRAS.324..877A}
{Allen}, S.~W., {Ettori}, S., \& {Fabian}, A.~C. 2001{\natexlab{a}}, \mnras,
  324, 877

\bibitem[{{Allen} et~al.(2004){Allen}, {Schmidt}, {Ebeling}, {Fabian}, \& {van
  Speybroeck}}]{2004MNRAS.353..457A}
{Allen}, S.~W., {Schmidt}, R.~W., {Ebeling}, H., {Fabian}, A.~C., \& {van
  Speybroeck}, L. 2004, \mnras, 353, 457

\bibitem[{{Allen} et~al.(2001{\natexlab{b}}){Allen}, {Schmidt} \&
  {Fabian}}]{2001MNRAS.328L..37A}
{Allen}, S.~W., {Schmidt}, R.~W., \& {Fabian}, A.~C. 2001{\natexlab{b}},
  \mnras, 328, L37

\bibitem[{{Allen} et~al.(2002){Allen}, {Schmidt} \&
  {Fabian}}]{2002MNRAS.334L..11A}
{Allen}, S.~W., {Schmidt}, R.~W., \& {Fabian}, A.~C. 2002, \mnras, 334, L11

\bibitem[{{Anders} \& {Grevesse}(1989)}]{1989GeCoA..53..197A}
{Anders}, E. \& {Grevesse}, N. 1989, \gca, 53, 197

\bibitem[{{Arabadjis} et~al.(2004){Arabadjis}, {Bautz} \&
  {Arabadjis}}]{2004ApJ...617..303A}
{Arabadjis}, J.~S., {Bautz}, M.~W., \& {Arabadjis}, G. 2004, \apj, 617, 303

\bibitem[{{Arnaud} \& {Evrard}(1999)}]{1999MNRAS.305..631A}
{Arnaud}, M. \& {Evrard}, A.~E. 1999, \mnras, 305, 631

\bibitem[{{Arnaud} et~al.(2005){Arnaud}, {Pointecouteau} \&
  {Pratt}}]{2005astro.ph..2210A}
{Arnaud}, M., {Pointecouteau}, E., \& {Pratt}, G.~W. 2005, \aap, submitted
  (astro-ph/0502210)

\bibitem[{{Bell} et~al.(2003){Bell}, {McIntosh}, {Katz}, \&
  {Weinberg}}]{2003ApJS..149..289B}
{Bell}, E.~F., {McIntosh}, D.~H., {Katz}, N., \& {Weinberg}, M.~D. 2003, \apjs,
  149, 289

\bibitem[{{Borgani} et~al.(2004){Borgani}, et~al.}]{2004MNRAS.348.1078B}
{Borgani}, S., et~al. 2004, \mnras, 348, 1078

\bibitem[{{Bryan} \& {Norman}(1998)}]{bryan_norman98}
{Bryan}, G.~L. \& {Norman}, M.~L. 1998, \apj, 495, 80

\bibitem[{{Bullock} et~al.(2001){Bullock}, {Kolatt}, {Sigad}, {Somerville},
  {Kravtsov}, {Klypin}, {Primack}, \& {Dekel}}]{bullock_etal01}
{Bullock}, J.~S., {Kolatt}, T.~S., {Sigad}, Y., {Somerville}, R.~S.,
  {Kravtsov}, A.~V., {Klypin}, A.~A., {Primack}, J.~R., \& {Dekel}, A. 2001,
  \mnras, 321, 559

\bibitem[{{Buote} \& {Lewis}(2004)}]{2004ApJ...604..116B}
{Buote}, D.~A. \& {Lewis}, A.~D. 2004, \apj, 604, 116

\bibitem[{{Cavaliere} \& {Fusco-Femiano}(1978)}]{1978A&A....70..677C}
{Cavaliere}, A. \& {Fusco-Femiano}, R. 1978, \aap, 70, 677

\bibitem[{{David} et~al.(2001){David}, {Nulsen}, {McNamara}, {Forman}, {Jones},
  {Ponman}, {Robertson}, \& {Wise}}]{2001ApJ...557..546D}
{David}, L.~P., {Nulsen}, P.~E.~J., {McNamara}, B.~R., {Forman}, W., {Jones},
  C., {Ponman}, T., {Robertson}, B., \& {Wise}, M. 2001, \apj, 557, 546

\bibitem[{{De Grandi} \& {Molendi}(2002)}]{2002ApJ...567..163D}
{De Grandi}, S. \& {Molendi}, S. 2002, \apj, 567, 163

\bibitem[{{Dolag} et~al.(2004{\natexlab{a}}){Dolag}, {Bartelmann}, {Perrotta},
  {Baccigalupi}, {Moscardini}, {Meneghetti}, \& {Tormen}}]{dolag_etal04}
{Dolag}, K., {Bartelmann}, M., {Perrotta}, F., {Baccigalupi}, C., {Moscardini},
  L., {Meneghetti}, M., \& {Tormen}, G. 2004{\natexlab{a}}, \aap, 416, 853

\bibitem[{{Dolag} et~al.(2004{\natexlab{b}}){Dolag}, {Jubelgas}, {Springel},
  {Borgani}, \& {Rasia}}]{2004ApJ...606L..97D}
{Dolag}, K., {Jubelgas}, M., {Springel}, V., {Borgani}, S., \& {Rasia}, E.
  2004{\natexlab{b}}, \apjl, 606, L97

\bibitem[{{Dubinski} \& {Carlberg}(1991)}]{dubinski_carlberg91}
{Dubinski}, J. \& {Carlberg}, R.~G. 1991, \apj, 378, 496

\bibitem[{{Eke} et~al.(1996){Eke}, {Cole} \& {Frenk}}]{eke_etal96}
{Eke}, V.~R., {Cole}, S., \& {Frenk}, C.~S. 1996, \mnras, 282, 263

\bibitem[{{Eke} et~al.(1998){Eke}, {Cole}, {Frenk}, \&
  {Henry}}]{1998MNRAS.298.1145E}
{Eke}, V.~R., {Cole}, S., {Frenk}, C.~S., \& {Henry}, J.~P. 1998, \mnras, 298,
  1145

\bibitem[{{Eke} et~al.(2001){Eke}, {Navarro} \& {Steinmetz}}]{eke_etal01}
{Eke}, V.~R., {Navarro}, J.~F., \& {Steinmetz}, M. 2001, \apj, 554, 114

\bibitem[{{Ettori}(2003)}]{2003MNRAS.344L..13E}
{Ettori}, S. 2003, \mnras, 344, L13

\bibitem[{{Ettori} \& {Fabian}(1999)}]{1999MNRAS.305..834E}
{Ettori}, S. \& {Fabian}, A.~C. 1999, \mnras, 305, 834

\bibitem[{{Evrard} et~al.(1996){Evrard}, {Metzler} \&
  {Navarro}}]{1996ApJ...469..494E}
{Evrard}, A.~E., {Metzler}, C.~A., \& {Navarro}, J.~F. 1996, \apj, 469, 494

\bibitem[{{Faltenbacher} et~al.(2005){Faltenbacher}, {Kravtsov}, {Nagai}, \&
  {Gottl{\" o}ber}}]{faltenbacher_etal05}
{Faltenbacher}, A., {Kravtsov}, A.~V., {Nagai}, D., \& {Gottl{\" o}ber}, S.
  2005, \mnras, 358, 139

\bibitem[{{Finoguenov} et~al.(2001){Finoguenov}, {Reiprich} \& {B{\"
  o}hringer}}]{2001A&A...368..749F}
{Finoguenov}, A., {Reiprich}, T.~H., \& {B{\" o}hringer}, H. 2001, \aap, 368,
  749

\bibitem[{{Frenk} et~al.(1990){Frenk}, {White}, {Efstathiou}, \&
  {Davis}}]{1990ApJ...351...10F}
{Frenk}, C.~S., {White}, S.~D.~M., {Efstathiou}, G., \& {Davis}, M. 1990, \apj,
  351, 10

\bibitem[{{Gnedin} et~al.(2004){Gnedin}, {Kravtsov}, {Klypin}, \&
  {Nagai}}]{2004ApJ...616...16G}
{Gnedin}, O.~Y., {Kravtsov}, A.~V., {Klypin}, A.~A., \& {Nagai}, D. 2004, \apj,
  616, 16

\bibitem[{{Helsdon} \& {Ponman}(2000)}]{2000MNRAS.315..356H}
{Helsdon}, S.~F. \& {Ponman}, T.~J. 2000, \mnras, 315, 356

\bibitem[{{Henry}(1997)}]{1997ApJ...489L...1H}
{Henry}, J.~P. 1997, \apjl, 489, L1

\bibitem[{{Henry}(2000)}]{2000ApJ...534..565H}
{Henry}, J.~P. 2000, \apj, 534, 565

\bibitem[{{Henry} \& {Arnaud}(1991)}]{1991ApJ...372..410H}
{Henry}, J.~P. \& {Arnaud}, K.~A. 1991, \apj, 372, 410

\bibitem[{{Horner} et~al.(1999){Horner}, {Mushotzky} \&
  {Scharf}}]{1999ApJ...520...78H}
{Horner}, D.~J., {Mushotzky}, R.~F., \& {Scharf}, C.~A. 1999, \apj, 520, 78

\bibitem[{{Hu} \& {Kravtsov}(2003)}]{2003ApJ...584..702H}
{Hu}, W. \& {Kravtsov}, A.~V. 2003, \apj, 584, 702

\bibitem[{{Ikebe} et~al.(2002){Ikebe}, {Reiprich}, {B{\" o}hringer}, {Tanaka},
  \& {Kitayama}}]{2002A&A...383..773I}
{Ikebe}, Y., {Reiprich}, T.~H., {B{\" o}hringer}, H., {Tanaka}, Y., \&
  {Kitayama}, T. 2002, \aap, 383, 773

\bibitem[{{Jenkins} et~al.(2001){Jenkins}, {Frenk}, {White}, {Colberg}, {Cole},
  {Evrard}, {Couchman}, \& {Yoshida}}]{2001MNRAS.321..372J}
{Jenkins}, A., {Frenk}, C.~S., {White}, S.~D.~M., {Colberg}, J.~M., {Cole}, S.,
  {Evrard}, A.~E., {Couchman}, H.~M.~P., \& {Yoshida}, N. 2001, \mnras, 321,
  372

\bibitem[{{Jing}(2000)}]{jing00}
{Jing}, Y.~P. 2000, \apj, 535, 30

\bibitem[{{Kay} et~al.(2004){Kay}, {Thomas}, {Jenkins}, \&
  {Pearce}}]{2004MNRAS.355.1091K}
{Kay}, S.~T., {Thomas}, P.~A., {Jenkins}, A., \& {Pearce}, F.~R. 2004, \mnras,
  355, 1091

\bibitem[{{Kazantzidis} et~al.(2004){Kazantzidis}, {Kravtsov}, {Zentner},
  {Allgood}, {Nagai}, \& {Moore}}]{kazantzidis_etal04}
{Kazantzidis}, S., {Kravtsov}, A.~V., {Zentner}, A.~R., {Allgood}, B., {Nagai},
  D., \& {Moore}, B. 2004, \apjl, 611, L73

\bibitem[{{Kravtsov} et~al.(2005){Kravtsov}, {Nagai} \&
  {Vikhlinin}}]{kravtsov_etal05}
{Kravtsov}, A.~V., {Nagai}, D., \& {Vikhlinin}, A.~A. 2005, \apj, 625, 588

\bibitem[{{Lewis} et~al.(2003){Lewis}, {Buote} \&
  {Stocke}}]{2003ApJ...586..135L}
{Lewis}, A.~D., {Buote}, D.~A., \& {Stocke}, J.~T. 2003, \apj, 586, 135

\bibitem[{{Lilje}(1992)}]{1992ApJ...386L..33L}
{Lilje}, P.~B. 1992, \apjl, 386, L33

\bibitem[{{Lin} \& {Mohr}(2004)}]{2004ApJ...617..879L}
{Lin}, Y. \& {Mohr}, J.~J. 2004, \apj, 617, 879

\bibitem[{{Lin} et~al.(2003){Lin}, {Mohr} \& {Stanford}}]{2003ApJ...591..749L}
{Lin}, Y., {Mohr}, J.~J., \& {Stanford}, S.~A. 2003, \apj, 591, 749

\bibitem[{{Markevitch}(1998)}]{1998ApJ...504...27M}
{Markevitch}, M. 1998, \apj, 504, 27

\bibitem[{{Markevitch} et~al.(1998){Markevitch}, {Forman}, {Sarazin}, \&
  {Vikhlinin}}]{1998ApJ...503...77M}
{Markevitch}, M., {Forman}, W.~R., {Sarazin}, C.~L., \& {Vikhlinin}, A. 1998,
  \apj, 503, 77

\bibitem[{{Markevitch} \& {Vikhlinin}(1997)}]{1997ApJ...491..467M}
{Markevitch}, M. \& {Vikhlinin}, A. 1997, \apj, 491, 467

\bibitem[{{Markevitch} et~al.(1999){Markevitch}, {Vikhlinin}, {Forman}, \&
  {Sarazin}}]{1999ApJ...527..545M}
{Markevitch}, M., {Vikhlinin}, A., {Forman}, W.~R., \& {Sarazin}, C.~L. 1999,
  \apj, 527, 545

\bibitem[{{Mathews}(1978)}]{1978ApJ...219..413M}
{Mathews}, W.~G. 1978, \apj, 219, 413

\bibitem[{{McNamara} et~al.(2005){McNamara}, {Nulsen}, {Wise}, {Rafferty},
  {Carilli}, {Sarazin}, \& {Blanton}}]{2005Natur.433...45M}
{McNamara}, B.~R., {Nulsen}, P.~E.~J., {Wise}, M.~W., {Rafferty}, D.~A.,
  {Carilli}, C., {Sarazin}, C.~L., \& {Blanton}, E.~L. 2005, \nat, 433, 45

\bibitem[{{Mohr} et~al.(1999){Mohr}, {Mathiesen} \&
  {Evrard}}]{1999ApJ...517..627M}
{Mohr}, J.~J., {Mathiesen}, B., \& {Evrard}, A.~E. 1999, \apj, 517, 627

\bibitem[{{Motl} et~al.(2004){Motl}, {Burns}, {Loken}, {Norman}, \&
  {Bryan}}]{2004ApJ...606..635M}
{Motl}, P.~M., {Burns}, J.~O., {Loken}, C., {Norman}, M.~L., \& {Bryan}, G.
  2004, \apj, 606, 635

\bibitem[{{Nagai} et~al.(2003){Nagai}, {Kravtsov} \&
  {Kosowsky}}]{2003ApJ...587..524N}
{Nagai}, D., {Kravtsov}, A.~V., \& {Kosowsky}, A. 2003, \apj, 587, 524

\bibitem[{{Navarro} et~al.(1996){Navarro}, {Frenk} \& {White}}]{navarro_etal96}
{Navarro}, J.~F., {Frenk}, C.~S., \& {White}, S.~D.~M. 1996, \apj, 462, 563

\bibitem[{{Navarro} et~al.(1997){Navarro}, {Frenk} \& {White}}]{navarro_etal97}
{Navarro}, J.~F., {Frenk}, C.~S., \& {White}, S.~D.~M. 1997, \apj, 490, 493

\bibitem[{{Neumann} \& {Arnaud}(1999)}]{1999A&A...348..711N}
{Neumann}, D.~M. \& {Arnaud}, M. 1999, \aap, 348, 711

\bibitem[{{Nevalainen} et~al.(2000){Nevalainen}, {Markevitch} \&
  {Forman}}]{2000ApJ...532..694N}
{Nevalainen}, J., {Markevitch}, M., \& {Forman}, W. 2000, \apj, 532, 694

\bibitem[{{Nulsen} et~al.(2005){Nulsen}, {Hambrick}, {McNamara}, {Rafferty},
  {Birzan}, {Wise}, \& {David}}]{2005ApJ...625L...9N}
{Nulsen}, P.~E.~J., {Hambrick}, D.~C., {McNamara}, B.~R., {Rafferty}, D.,
  {Birzan}, L., {Wise}, M.~W., \& {David}, L.~P. 2005, \apjl, 625, L9

\bibitem[{{Oukbir} \& {Blanchard}(1992)}]{1992A&A...262L..21O}
{Oukbir}, J. \& {Blanchard}, A. 1992, \aap, 262, L21

\bibitem[{{Pointecouteau} et~al.(2004){Pointecouteau}, {Arnaud}, {Kaastra}, \&
  {de Plaa}}]{2004A&A...423...33P}
{Pointecouteau}, E., {Arnaud}, M., {Kaastra}, J., \& {de Plaa}, J. 2004, \aap,
  423, 33

\bibitem[{{Pointecouteau} et~al.(2005){Pointecouteau}, {Arnaud} \&
  {Pratt}}]{2005A&A...435....1P}
{Pointecouteau}, E., {Arnaud}, M., \& {Pratt}, G.~W. 2005, \aap, 435, 1

\bibitem[{{Pratt} \& {Arnaud}(2002)}]{2002A&A...394..375P}
{Pratt}, G.~W. \& {Arnaud}, M. 2002, \aap, 394, 375

\bibitem[{{Press} et~al.(1992){Press}, {Teukolsky}, {Vettering}, \&
  {Flannery}}]{numerical_recipes}
{Press}, W.~H., {Teukolsky}, S.~A., {Vettering}, W.~T., \& {Flannery}, B.~P.,
  Numerical Recipes (Cambridge: Cambridge Univ.~Press, 1992)

\bibitem[{{Randall} et~al.(2002){Randall}, {Sarazin} \&
  {Ricker}}]{2002ApJ...577..579R}
{Randall}, S.~W., {Sarazin}, C.~L., \& {Ricker}, P.~M. 2002, \apj, 577, 579

\bibitem[{{Rasia} et~al.(2005){Rasia}, {Mazzotta}, {Borgani}, {Moscardini},
  {Dolag}, {Tormen}, {Diaferio}, \& {Murante}}]{2005ApJ...618L...1R}
{Rasia}, E., {Mazzotta}, P., {Borgani}, S., {Moscardini}, L., {Dolag}, K.,
  {Tormen}, G., {Diaferio}, A., \& {Murante}, G. 2005, \apjl, 618, L1

\bibitem[{{Rasia} et~al.(2004){Rasia}, {Tormen} \&
  {Moscardini}}]{2004MNRAS.351..237R}
{Rasia}, E., {Tormen}, G., \& {Moscardini}, L. 2004, \mnras, 351, 237

\bibitem[{{Readhead} et~al.(2004){Readhead}, et~al.}]{2004Sci...306..836R}
{Readhead}, A.~C.~S., et~al. 2004, Science, 306, 836

\bibitem[{{Rines} et~al.(1999){Rines}, {Forman}, {Pen}, {Jones}, \&
  {Burg}}]{1999ApJ...517...70R}
{Rines}, K., {Forman}, W., {Pen}, U., {Jones}, C., \& {Burg}, R. 1999, \apj,
  517, 70

\bibitem[{{Sanderson} \& {Ponman}(2003)}]{2003MNRAS.345.1241S}
{Sanderson}, A.~J.~R. \& {Ponman}, T.~J. 2003, \mnras, 345, 1241

\bibitem[{{Sanderson} et~al.(2003){Sanderson}, {Ponman}, {Finoguenov},
  {Lloyd-Davies}, \& {Markevitch}}]{2003MNRAS.340..989S}
{Sanderson}, A.~J.~R., {Ponman}, T.~J., {Finoguenov}, A., {Lloyd-Davies},
  E.~J., \& {Markevitch}, M. 2003, \mnras, 340, 989

\bibitem[{{Sarazin}(1988)}]{sarazin88}
{Sarazin}, C.~L., X-ray Emission from Clusters of Galaxies (Cambridge:
  Cambridge University Press, 1988)

\bibitem[{{Sheth} \& {Tormen}(1999)}]{1999MNRAS.308..119S}
{Sheth}, R.~K. \& {Tormen}, G. 1999, \mnras, 308, 119

\bibitem[{{Snowden} et~al.(1994){Snowden}, {McCammon}, {Burrows}, \&
  {Mendenhall}}]{1994ApJ...424..714S}
{Snowden}, S.~L., {McCammon}, D., {Burrows}, D.~N., \& {Mendenhall}, J.~A.
  1994, \apj, 424, 714

\bibitem[{{Spergel} et~al.(2003){Spergel}, et~al.}]{2003ApJS..148..175S}
{Spergel}, D.~N., et~al. 2003, \apjs, 148, 175

\bibitem[{{Sunyaev} \& {Zeldovich}(1972)}]{1972CoASP...4..173S}
{Sunyaev}, R.~A. \& {Zeldovich}, Y.~B. 1972, Comments on Astrophysics and Space
  Physics, 4, 173

\bibitem[{{Tasitsiomi} et~al.(2004){Tasitsiomi}, {Kravtsov}, {Gottl{\" o}ber},
  \& {Klypin}}]{tasitsiomi_etal04}
{Tasitsiomi}, A., {Kravtsov}, A.~V., {Gottl{\" o}ber}, S., \& {Klypin}, A.~A.
  2004, \apj, 607, 125

\bibitem[{{Vikhlinin}(2005)}]{astro-ph/0504098}
{Vikhlinin}, A. 2005, \apj{} submitted (astro-ph/0504098)

\bibitem[{{Vikhlinin} et~al.(1999){Vikhlinin}, {Forman} \&
  {Jones}}]{1999ApJ...525...47V}
{Vikhlinin}, A., {Forman}, W., \& {Jones}, C. 1999, \apj, 525, 47

\bibitem[{{Vikhlinin} et~al.(2005){Vikhlinin}, {Markevitch}, {Murray}, {Jones},
  {Forman}, \& {Van Speybroeck}}]{2004astro.ph.12306V}
{Vikhlinin}, A., {Markevitch}, M., {Murray}, S.~S., {Jones}, C., {Forman}, W.,
  \& {Van Speybroeck}, L. 2005, \apj{}, in press, astro-ph/0412306 (Paper~I)

\bibitem[{{Vikhlinin} et~al.(1998){Vikhlinin}, {McNamara}, {Forman}, {Jones},
  {Quintana}, \& {Hornstrup}}]{1998ApJ...502..558V}
{Vikhlinin}, A., {McNamara}, B.~R., {Forman}, W., {Jones}, C., {Quintana}, H.,
  \& {Hornstrup}, A. 1998, \apj, 502, 558

\bibitem[{{Voevodkin} \& {Vikhlinin}(2004)}]{2004ApJ...601..610V}
{Voevodkin}, A. \& {Vikhlinin}, A. 2004, \apj, 601, 610

\bibitem[{{Voit}(2005)}]{2005RvMP...77..207V}
{Voit}, G.~M. 2005, Reviews of Modern Physics, 77, 207

\bibitem[{{Wechsler} et~al.(2002){Wechsler}, {Bullock}, {Primack}, {Kravtsov},
  \& {Dekel}}]{wechsler_etal02}
{Wechsler}, R.~H., {Bullock}, J.~S., {Primack}, J.~R., {Kravtsov}, A.~V., \&
  {Dekel}, A. 2002, \apj, 568, 52

\bibitem[{{White}(2001)}]{2001A&A...367...27W}
{White}, M. 2001, \aap, 367, 27

\bibitem[{{White} et~al.(1993{\natexlab{a}}){White}, {Efstathiou} \&
  {Frenk}}]{1993MNRAS.262.1023W}
{White}, S.~D.~M., {Efstathiou}, G., \& {Frenk}, C.~S. 1993{\natexlab{a}},
  \mnras, 262, 1023

\bibitem[{{White} et~al.(1993{\natexlab{b}}){White}, {Navarro}, {Evrard}, \&
  {Frenk}}]{1993Natur.366..429W}
{White}, S.~D.~M., {Navarro}, J.~F., {Evrard}, A.~E., \& {Frenk}, C.~S.
  1993{\natexlab{b}}, \nat, 366, 429

\bibitem[{{Xu} et~al.(2001){Xu}, {Jin} \& {Wu}}]{2001ApJ...553...78X}
{Xu}, H., {Jin}, G., \& {Wu}, X. 2001, \apj, 553, 78

\end{thebibliography}

\begin{appendix}
  
\section{Comparison with previous determinations of the $M-T$ relation}
\label{sec:MT:comparisons}

The main differences of our analysis with the previous work on cluster
masses is in the more accurate measurements of the gas temperature and
density gradients at large radii. Comparison between different studies
is facilitated by formulation of the the hydrostatic equlibrium equation
([\ref{eq:hydro}]) using effective slopes of the density and temperature
profiles. 

Let us define the effective bas density slope as $\beta_{\text{eff}} =
-\frac{1}{3}\, d\log \rho/d\log r$; for the $\beta$-model,
$\beta_{\text{eff}}=\beta \,(r/r_c)^2/(1+(r/r_c)^2)\rightarrow\beta$ at
large radii. The equivalent quantity for the temperature profile is
$\beta_t = -1/3\, d\log T/d\log r$. For polytropic parameterization of
the temperature profile, $\beta_{\text{eff}}$ and $\beta_t$ are related
via
\begin{equation}\label{eq:betaT:gamma}
  \frac{d\log T}{d\log r} = (\gamma-1) \frac{d\log T}{d\log r} \quad
  \text{or} \quad \beta_t = (\gamma-1)\,\beta{_\text{eff}}
\end{equation}

The hydrostatic equilibrium equation~[\ref{eq:hydro}] can now be
rewritten as $M(r)\propto r\times T(r)\times
(\beta_{\text{eff}}+\beta_t)$. To compute the overdensity mass, we solve
equation of the type $M(r_\Delta)/r_\Delta^3 = C$, or
$r_\Delta^{-2}\,T(r)\,(\beta_{\text{eff}}+\beta_t)=C$.  Therefore, the
overdensity mass estimate scales as
\begin{equation}\label{eq:M:beta:scale}
  M_\Delta \propto T_0^{3/2}\,(T(r)/T_0)^{3/2}(\beta_{\text{eff}}+\beta_t)^{3/2},
\end{equation}
where $T_0$ is an average temperature, and normalization of the $M-T$
relation scales as
\begin{equation}\label{eq:MT:beta:scale}
  A = M_\Delta/T_0^{3/2} \propto (T(r)/T_0)^{3/2}(\beta_{\text{eff}}+\beta_t)^{3/2}
\end{equation}
The quantities $\beta_{\text{eff}}$ and $\beta_t$ derived for our
clusters at $r_{500}$ are shown in Fig.\ref{fig:slopes500}. 

\subsection{\emph{ASCA} Measurements by Nevalainen et al.{} and
  Finoguenov et al.{}}

% Jukka, r500, at 5 keV: 2.16e14 h**-1
% Finoguenov:            2.32e14 h**-1

\begin{figure}
  \vspace*{-2.5\baselineskip}
  \centerline{\includegraphics[width=0.478\linewidth]{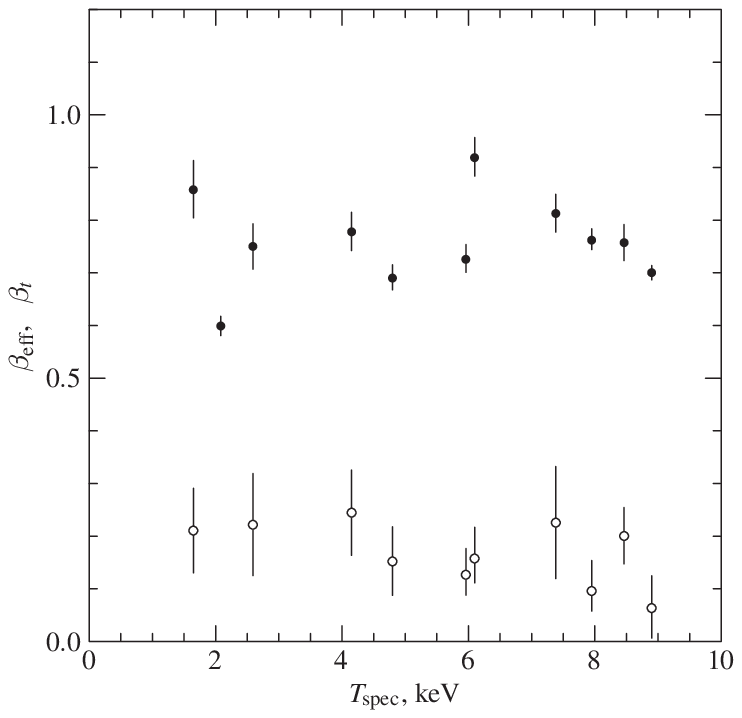}}
  \vspace*{-1\baselineskip}
  \caption{Effective slopes of gas density and temperature profiles at
    $r_{500}$}
  \label{fig:slopes500}
\end{figure}

\citet{2000ApJ...532..694N} and \citet{2001A&A...368..749F} use
\emph{ASCA} temperature profiles which agree with our \emph{Chandra}
measurements. However, their normalizations for the $M_{500}-T$ relation
for $T=5$~keV clusters are lower --- $2.2\times10^{14}\,h^{-1}\,M_\odot$
and $2.3\times10^{14}\,h^{-1}\,M_\odot$ in
\citeauthor{2000ApJ...532..694N} and \citeauthor{2001A&A...368..749F},
respectively, compared with our value of
$2.93\times10^{14}\,h^{-1}\,M_\odot$ (Table~\ref{tab:mtfit}, for
$T_{\text{spec}}$). There are subtle differences between these studies
in how the average temperatures are defined and how the measurements are
scaled to $z=0$. The most significant effect, however, is steepening of
the gas density profile at large radii that is clearly present in
high-quality \emph{Chandra} data and was hard to detect with earlier
X-ray telescopes. This steepening was effectively missed by
\citeauthor{2000ApJ...532..694N} and \citeauthor{2001A&A...368..749F}
because they used pure $\beta$-model fits determined mainly by the data
in the inner regions. Indeed, the average $\beta_{\text{eff}}$ at
$r_{500}$ is $0.78$ for our clusters (Fig.\,\ref{fig:slopes500}), while
it is $\approx 0.65$ for $T>3$~keV clusters in
\citeauthor{2001A&A...368..749F} (see their Fig.\,5). The average
$\beta_t$ for our clusters is 0.17, while the
\citeauthor{2001A&A...368..749F} value is 0.11 from their average
polytropic index of $\gamma\approx1.17$ for hot clusters (see
eq.\,[\ref{eq:betaT:gamma}]). These differences in the slopes should
lead (eq.\,[\ref{eq:M:beta:scale}]) to a mismatch in mass estimates by a
factor of $(0.78+0.15)^{3/2}/(0.65+0.11)^{3/2}\simeq 1.35$, explaining
the offset between our $M-T$ normalizations. We note here that
undetected steepening of the gas density profiles at large radii was
suggested by \citet{2004MNRAS.348.1078B} as a possible reason for low
normalization of the \citeauthor{2001A&A...368..749F} $M-T$ relation. 

We find no detectable trends in either $\beta_{\text{eff}}$ or $\beta_t$
with cluster temperature (Fig.\,\ref{fig:slopes500}). The gas density
profiles of cool clusters are indeed flatter in the inner region but
they steepen significantly at $r\sim 0.5\,r_{500}$
(Fig.\,\ref{fig:scaled:profs}). This absence of trends in
$\beta_{\text{eff}}$ and $\beta_t$ is in fact the main reason why our
slopes of mass-temperature relation are close to $1.5$
(cf.~eq.\,[\ref{eq:M:beta:scale}]). In contrast, earlier studies based
on $\beta$-model fits to the X-ray brightness profiles consistently
found very flat density slopes for 1--2~keV clusters. For example, the
average $\beta_{\text{eff}}$ for such clusters is $\simeq 0.5$ in
\citeauthor{2001A&A...368..749F}, while for $T\sim10$~keV clusters, they
find $\beta_{\text{eff}}\approx 0.7$. It follows from
eq.\,[\ref{eq:M:beta:scale}] that such a trend in $\beta_{\text{eff}}$
should steepen the slope of the $M-T$ relation from $\alpha=1.5$ to
$\alpha \sim 1.72$.

\subsection{XMM-Newton Measurements of Arnaud et al.{}}
\label{sec:comparison:arnaud.etal}

Our normalization of the $M-T$ relation is very close to that derived by
\citet{2005astro.ph..2210A} using \emph{XMM-Newton} data. Given the
systematic difference in the temperature profiles at large radii between
our two studies (see the discussion in \S\,\ref{sec:T:prof}), this is
not expected and merits a clarification. In the Arnaud et al.{} works,
temperature profiles are generally interpreted to be consistent with
constant at large radii \citep{2005A&A...435....1P}.  Therefore, the
quantities to use in eqs.\ (A2-A3), $\beta_t\simeq0$ and $T(r)/T_0\simeq
1$ near $r_{500}$ for the \emph{XMM-Newton} masses.  E.~Pointecouteau
and M.~Arnaud kindly provided the average gas density slope for their
sample, $\langle\beta_{\text{eff}}\rangle=0.71$, only slightly below our
value.  For our sample, the relevant quantities at $r_{500}$ are
$\langle\beta_{\text{eff}}\rangle=0.78$, $\langle\beta_t\rangle=0.17$,
and $\langle T(r)/T_0\rangle=0.67$, where $T_0$ is the spectroscopic
average temperature (see \S~\ref{sec:mean:T}). From
eq.\,[\ref{eq:MT:beta:scale}] we would then expect the \emph{XMM-Newton}
$M_{500}-T$ normalization to be a factor of $1.18$ higher than our
value, while in fact it is slightly lower (Fig.\ref{fig:MT:Tspec}).

The reason is that for the $M-T$ relation, Arnaud et al.{} used cluster
masses derived by fitting an NFW model to the data within $\sim r_{1000}$,
their maximum radius of observation. This gives systematically lower masses
at $r\gtrsim r_{500}$ than the values obtained by direct hydrostatic
derivation using extrapolation of the isothermal temperature profiles
(E.~Pointecouteau \& M.~Arnaud, private communication). In effect, the
masses from the NFW fit imply declining temperature profiles at large radii,
such as those observed by \emph{Chandra}. To summarize, the agreement
between our $M-T$ normalizations is somewhat a coincidence.

Note also that the \emph{XMM-Newton} cluster temperatures are systematically
lower than those from \emph{Chandra} \citep[see
e.g.,][]{2004astro.ph.12306V}), but this does not change the normalization
of the $M-T$ relation as it moves the clusters along the relation.

We also used the results from the \citet{2005A&A...435....1P} sample to
check our mass derivation algorithm. The published \emph{XMM-Newton}
temperature and density profiles were used as input to our procedure. 
The obtained mass profiles were nearly identical to those from
\citeauthor{2005A&A...435....1P}

\section{Comparison with \fgas{} measurements of Allen et al.} 
\label{sec:comp:allen}

Gas fractions within $r_{2500}$ derived from \emph{Chandra} observations
of relaxed clusters were recently used for cosmological constraints in a
series of papers by \citet[and references therein]{2004MNRAS.353..457A}. 
There are significant differences in the $f_{\text{gas},2500}$ derived
in our analysis and in
\citeauthor{2004MNRAS.353..457A}, as outlined below. 

For all our $T>5$~keV clusters except A2390, we derive lower values of
\fgas{}. Our average value for these clusters is $\fgas=0.091\pm0.002$,
compared to an average of $0.117\pm0.002$ in
\citeauthor{2004MNRAS.353..457A} (same temperature range and same
cosmology). The same $\sim 25\%$ difference holds for the four clusters
common in both samples, A2029, A478, A1413, and A383 (see Table~2 in
\citealt{2004MNRAS.353..457A} and our Table~\ref{tab:masses}). 

We also do not observe flattening of $\fgas(r)$ at radii within
$r_{2500}$ as reported, e.g., in \citet*{2002MNRAS.334L..11A}. Note,
however, that a larger sample presented in \citet{2004MNRAS.353..457A}
shows more variety in the behavior of $\fgas(r)$ profiles. 

Profiles for individual clusters in the \citeauthor{2004MNRAS.353..457A}
sample were either not published or published prior to significant
\emph{Chandra} calibration updates. Therefore, we are unable to perform
a more detailed object-to-object comparison. 

\end{appendix}

\end{document}